\documentclass[twocolumn,showkeys,superscriptaddress]{revtex4}
\usepackage{graphicx}
\usepackage{bm}
\usepackage{color}
\usepackage{amsmath}
\usepackage{amsfonts}
\usepackage{amssymb}
\usepackage{natbib}
\usepackage{latexsym}
\usepackage{rotating}
\usepackage{lscape}
\usepackage{url}

\begin{document}

\title{Phenomenological dynamics of COVID--19 pandemic: meta--analysis for adjustment parameters }

\author{Sergio A. Hojman}
\email{sergio.hojman@uai.cl}
\affiliation{Departamento de Ciencias, Facultad de Artes Liberales,
Universidad Adolfo Ib\'a\~nez, Santiago 7491169, Chile.}
\affiliation{Departamento de F\'{\i}sica, Facultad de Ciencias, Universidad de Chile,
Santiago 7800003, Chile.}
\affiliation{Centro de Recursos Educativos Avanzados,
CREA, Santiago 7500018, Chile.}
\author{Felipe A. Asenjo}
\email{felipe.asenjo@uai.cl}
\affiliation{Facultad de Ingenier\'ia y Ciencias,
Universidad Adolfo Ib\'a\~nez, Santiago 7491169, Chile.}


\begin{abstract}
We present a  phenomenological procedure of dealing with the COVID--19 data provided by government health agencies of eleven different countries. Instead of using the (exact or approximate) solutions to the SIR (or other) model(s) to fit the data by adjusting the time--independent parameters included in those models, we introduce dynamical parameters whose time--dependence may be phenomenologically obtained by adequately extrapolating  a chosen subset of the daily provided data. This phenomenological approach works extremely well to properly adjust the number of infected (and removed) individuals in time,  for the  countries we consider. Besides, it can handle the sub--epidemic events that some countries may experience.
In this way, we obtain the evolution of the pandemic without using any {\it a priori} model based on differential equations.
\end{abstract}


\maketitle

\section{Introduction}

Pandemic propagation models are usually described by systems of first order ordinary non--linear coupled differential equations such as it is the case for the well--known SIS and SIR models, for instance. Numerous classical as well as very recent articles have been written to deal with this problem \cite{sir,abd,bai,het,katia,har,boh,wuhan1,harko}. The dynamical variables in such models are usually denoted by $S(t)$, $I(t)$ and $R(t)$, which are functions of a single time variable $t$ and denote the number of susceptible individuals (which may get infected), the number of infected individuals and the number of recovered or removed individuals (which some time after becoming infected are either immune or dead), respectively.

In addition to the dynamical variables, the models introduce time--independent parameters (usually denoted by greek letters) which describe the intensity of the coupling between the  variables. These parameters clearly depend on the  behavior of quarantined people in different countries which, in turn depend on the containment policies implemented by different governments in different countries.

It is not difficult to realize when trying to fit the data informed by health governmental institutions of different countries that they cannot be fitted by solutions of the model's equations for time--independent parameters, due to the fact that the policies change over time and so does the behavior of the societies.
Strictly speaking, the so--called parameters are really dynamical variables whose time evolution equations are difficult to hypothesize 
or construct due to the fact that their time evolution depends on people idiosyncrasy and government policies, which are almost impossible to foresee.  In order to overcome this issue, some (or all) of the parameters in those pandemic models can be promoted to time--dependent variables (see, for example, Refs.~\cite{zhang,lorenzo,Mummert1,Mummert2,capis,Pollicott,hadeler,Ungarala,khan,chowell,chowell2,chowell3}). However, the exact time dependence  is still unknown in those new variables, and the evolution of the pandemic is still very sensitive to the used model.

Because of these difficulties, in this work we propose an analysis on the study of the daily change of such parameters in order to retrieve the dynamical information of a pandemic. 
The estimation of the time--dependence behavior of the parameters of epidemiological models
have been an active research field in the past \cite{zhang,lorenzo,Mummert1,Mummert2,capis,Pollicott,hadeler,Ungarala,khan,chowell,chowell2,chowell3}. 
However, our procedure is different to previous ones, as it does not require any set of differential equations as a model. 
Instead of studying the total data set that gives origin to total structure of the pandemic, we focus in the study of time evolution of the parameters that produce such total structure, i.e., we do a meta--analysis of the data set system.
Therefore, rather than solving a model described by a set of differential equations with solutions that fit the data, a meta--analysis proceeds in the opposite direction, finding the global  evolution of the parameters, and thus obtaining a model. This procedure allows us to find the global dynamical behavior of data by studying the day-to-day evolution of the adjustment parameters. In this way, we can extract the time-dependent information of the system without solving any differential equation set. Hence, we are able to solve the system in a phenomenological fashion. This proper meta--analysis of the adjustment parameters can provide the kind of information that is needed in order to have a better understanding of the evolution of the pandemic.
The main goal of this work is to show that this procedure gives us global information of the changes in the spread of the disease.

The adjustment parameters in this meta--analysis, hereafter called meta--parameters, are no longer considered as constants, and they can be extracted directly from the data information. It is the purpose of this work to delineate a systematic procedure to estimate the dynamics of meta--parameters. We show how these meta--parameters substantially improve the understanding of the global evolution of the pandemic. This is exemplified for the case of actual current data from eleven  countries. These are Italy, United States, Canada, United Kingdom, Spain, Poland, Austria, Germany, Portugal, New Zealand and France.

In order to put  our proposal in context, let us consider first the SIR model as an example for pandemic evolution. This model considers three kind of populations, the susceptible $S=S(t)$, the infected $I=I(t)$ and the recovered $R=R(t)$ population, respectively, all of them evolving in time. Besides, the total population $N=S+I+R$, is constant in time.
The three variables 
 are related by the differential system $\dot S=-\alpha S I$, $\dot I=\alpha S I-\beta I$, and $\dot R=\beta I$ (where $\dot{A}(t) \equiv d A(t)/dt$). Here, $\alpha$ and $\beta$ are constant parameters that contain the relevant information for pandemic evolution. Our lack of knowledge on how the pandemic evolves is hidden in such parameters.
Although no explicit exact solution for $R=R(t)$
 is known, it is straightforward to show that at second order in an expansion around $\alpha R/\beta$, we can obtain the solution  for the recovered population as a function of time \cite{tesis,bailey}, given by $R(t)\approx  r_1\tanh (r_2\, t-r_3)+r_4$.
Here, 
$r_1=\gamma \beta^2/(\alpha^2 S_0)$, $r_2=\beta\gamma/2$, $r_3=\tanh^{-1}\left(\alpha S_0/(\beta\gamma)-1/\gamma\right)$, and $r_4=\beta/\alpha-\beta^2/(\alpha^2S_0)$, are all constants,
in terms of $\gamma=\left[(\alpha S_0/\beta-1)^2+2 S_0 I_0\alpha^2/\beta^2 \right]^{1/2}$, where $S_0$ and $I_0$ are the initial values of susceptible and infected
 populations at $t=0$. It has been assumed that the initial value of the recovered population $R_0=0$. 
It is important to realize that the system is now completely solved, as the infected population 
can be 
readily obtained by
$I(t)={\dot R}/{\beta}\approx ({r_1\gamma}/{2}){\mbox{sech}}^2(r_2\, t-r_3)$,
while the susceptible population is $S(t)=N-I(t)-R(t)$.
Those solutions are often used to study in an approximated manner the pandemic evolution. However, they fail to describe correctly its dynamics when social conditions change or different governmental decisions are taken along the progress of the pandemic.

In the following sections we show how better fitting results can be achieved by the procedure of using a hyperbolic tangent function  to fit  the data  of recovered population during the pandemic, as a starting point for the procedure  using meta--parameters. The evolution of these meta--parameters is obtained by analyzing  day-by-day the same data that they adjust. This procedure 
gives a precise figure of the increment of infected individuals. Therefore, the meta--analysis produces a better fitting of the estimation of the temporal behavior of infected population, thus solving the pandemic dynamics in a phenomenological way.
The final solution obtained from the data fitting procedure will not be a solution of the SIR model, neither of any other simple model described by first-order differential equations.


\section{Phenomenological treatment for pandemic dynamics}

In this section we describe the phenomenological procedure to estimate the evolution of the infected population. 

We start with the data set $R_j$ for the recovered population   at day $j$,  with $j=1,...,N$  measuring lapsed days, with a final day $N$. 
The information of this data set is equivalent to the cumulative integral or sum of infected cases. The data are obtained from Ref.~\cite{oneworld}. By using the information in $R_j$, we can infer the infected population data as 
\begin{equation}
I_{j}=R_{j}-R_{j-1}\, ,
\label{infecteddataresta}
\end{equation}
 at day $j$. For this work, we have used data of Ref.~\cite{oneworld} until June 19th, 2020.

For such given data set for the recovered population, a global dynamical behavior can be found by fitting the curve $R_j\rightarrow R(t)$, where now the continuous recovered population function is given by
\begin{equation}
R(t)= a\left(\tanh \left(b\, t+c\right)- \tanh c\right)+R_0\, ,
\label{Req1b}
\end{equation}
where $a$, $b$ and $c$ are  global constant adjustment parameters,
and we have assumed that 
 the relevant data to perform any analysis start with  $R_0\neq 0$, by properly setting the initial time $t=0$ of our analysis. By global we refer that the adjustment is for the total lapsed time $N$.
Notice that recovered population curve \eqref{Req1b} is not equal to the approximated solution emerging from  the SIR system. We show below that \eqref{Req1b} is a good global fitting for the recovered population. 
On the other hand, the global infected population dynamics is assumed to be found as $I_j\rightarrow I(t)=\dot R(t)$, which gives 
\begin{equation}
I(t)= a\, b\, {\mbox{sech}}^2\left(b\, t+c\right)\, ,
\label{Ieq1}
\end{equation}

Now, let us perform the meta--analysis of fitting \eqref{Req1b} for the recovered population. Let us promote the three parameters used in \eqref{Req1b} to meta--parameters $a\rightarrow a_1(t)$, $b\rightarrow a_2(t)$, and $c\rightarrow a_3(t)$. These meta--parameters are no longer global constants. Their dynamics must be obtained considering the new information that brings any new day that it is added to the data set of recovered population. 
For each time $j$ ($j=1,...,N$), the meta--parameters $a_i$  ($i=1,2,3$)
are found by fitting the curve \eqref{Req1b} to the data, by using them as constant adjustment parameters for such time.
 As the amount of data grows with time, the value of each meta--parameter varies, taking into account the different behaviors that the governments or the society may have at different times.  
After several iterations are performed for different times and fitting curves \eqref{Req1b}, a regular and dynamical behavior of each meta--parameter starts to emerge. 
This regularity starts at some time $\tau$ for the three meta--parameters, and it depends on each particular studied case.
All of this implies that meta--parameters are not globally constant in time, and now their global time--dependence  $a_i=a_i(t)$ is apparent, and more important, it can be deduced.

Once this stage is reached, the complete dynamical solution for each meta--parameter is established, and 
 solution \eqref{Req1b} for the recovered population
can now be promoted to the function
\begin{equation}\label{Req2}
R_M(t)= a_1(t)\left[ \tanh \left(a_2(t)\, t-a_3(t)\right)- \tanh\left(a_3(t)\right) \right]-R_0\, ,
\end{equation}
which produces a dramatic departure from solution \eqref{Req1b} itself.

With all the above, the new meta--parameter fitting function \eqref{Req2} 
contains more precise information of the daily changes of the recovered population compared with the fitting function \eqref{Req1b}. In other words, its derivative represents a more accurate description of the infected population curve, which can be calculated as
\begin{equation}
I_M(t)=\dot R_M(t)\, ,
\label{Ieq2}
\end{equation}
which anew, turns out to be different from function \eqref{Ieq1}.

It is shown below, that when meta--parameters have achieved a regular dynamics, they all can be described in the form ($i=1,2,3$)
\begin{equation}
a_i(t)\approx \sum_{k=0}^Z a_{ki} t^k\, ,
\label{formmetaparametera}
\end{equation}
with constant coefficients $a_{ki}$, and $Z>0$. Several meta--analysis produce very good agreement with the data for $Z=2$. On the other hand, and remarkably, 
to describe the sub--epidemic behavior of different countries, it is enough to consider $Z=4$. This shows that any change in the information of the data evolves in an ordered way and it can be recovered through the study of the meta--parameters.

In order to quantify how both infected population fittings
differ from each other, we define the global adjustment function $\epsilon=E_M/E_C$, as the ratio between $E_M={\sum_{j=2}^N \left[I_M(j)-I_j\right]^2}$, which is the error function 
for the meta--analysis of infected population \eqref{Ieq2}, and $E_C={\sum_{j=2}^N \left[I(j)-I_j\right]^2}$, that is
the error function 
for the  fitting of infected population \eqref{Ieq1} with constant adjustment parameters.
The case of $\epsilon<1$ implies a better fitting curve for the infected population dynamics
 due to the meta--analysis.

Below, we present examples for different countries that explicitly show the strength of this phenomenological dynamical analysis, and its better fitting to the existent data.
It is the goal of this work to search for the explicit form of each meta--parameter for the studied countries, finding in this way the underlying dynamical structure of their pandemic scenarios, and thus determine $R_M$ and $I_M$.
Besides, we show that this phenomenological model can also take into consideration the sub--epidemics ocurring during the whole time lapsed for the pandemic in some countries. The sub--epidemic behavior is coded in the  time dependence of meta--parameters.

With all this in mind, let us discuss the pandemic data
for cumulative infectious cases as evolving in time for eleven different countries, and how the phenomenological dynamical procedure applies to each of them.
We use the case of Italy to carefully explain each step in the procedure, as it is straightforwardly replicated for the other country cases.

\begin{figure*}[ht]
\begin{tabular}{cc}
  \includegraphics[width=80mm]{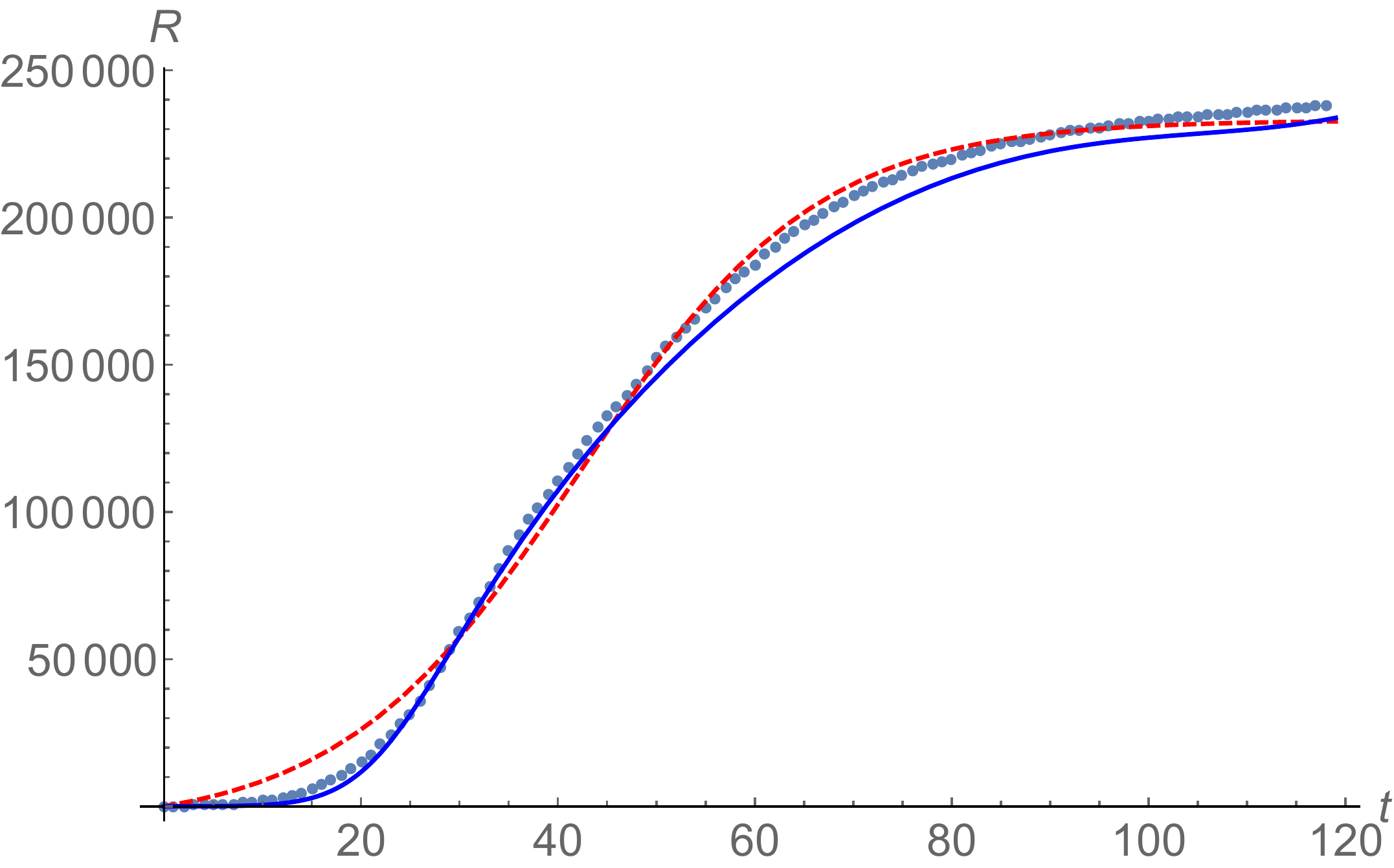} &   \includegraphics[width=80mm]{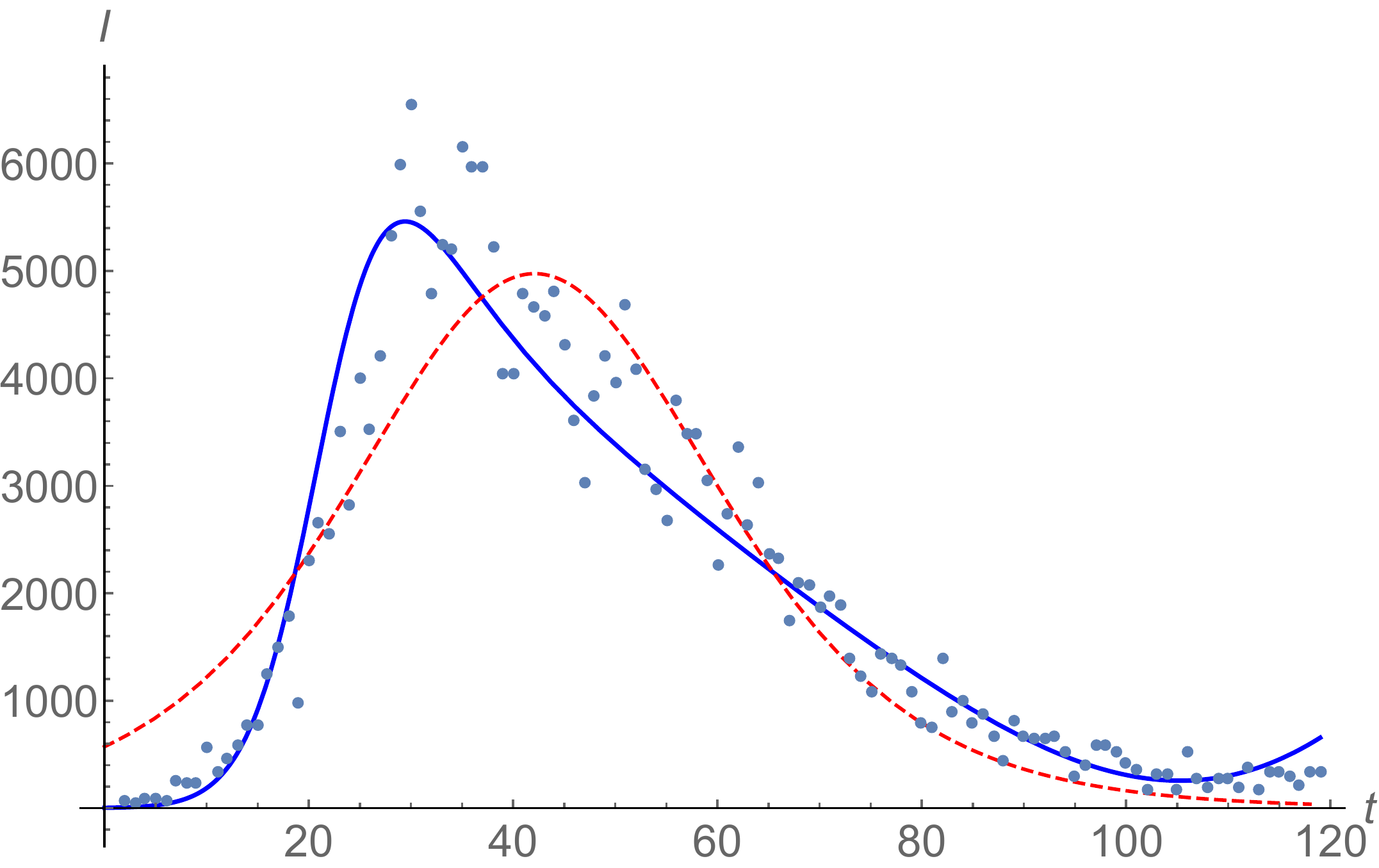} \\
(a) Recovered population for Italy & (b)  Infected population for Italy \\[6pt]
 \includegraphics[width=80mm]{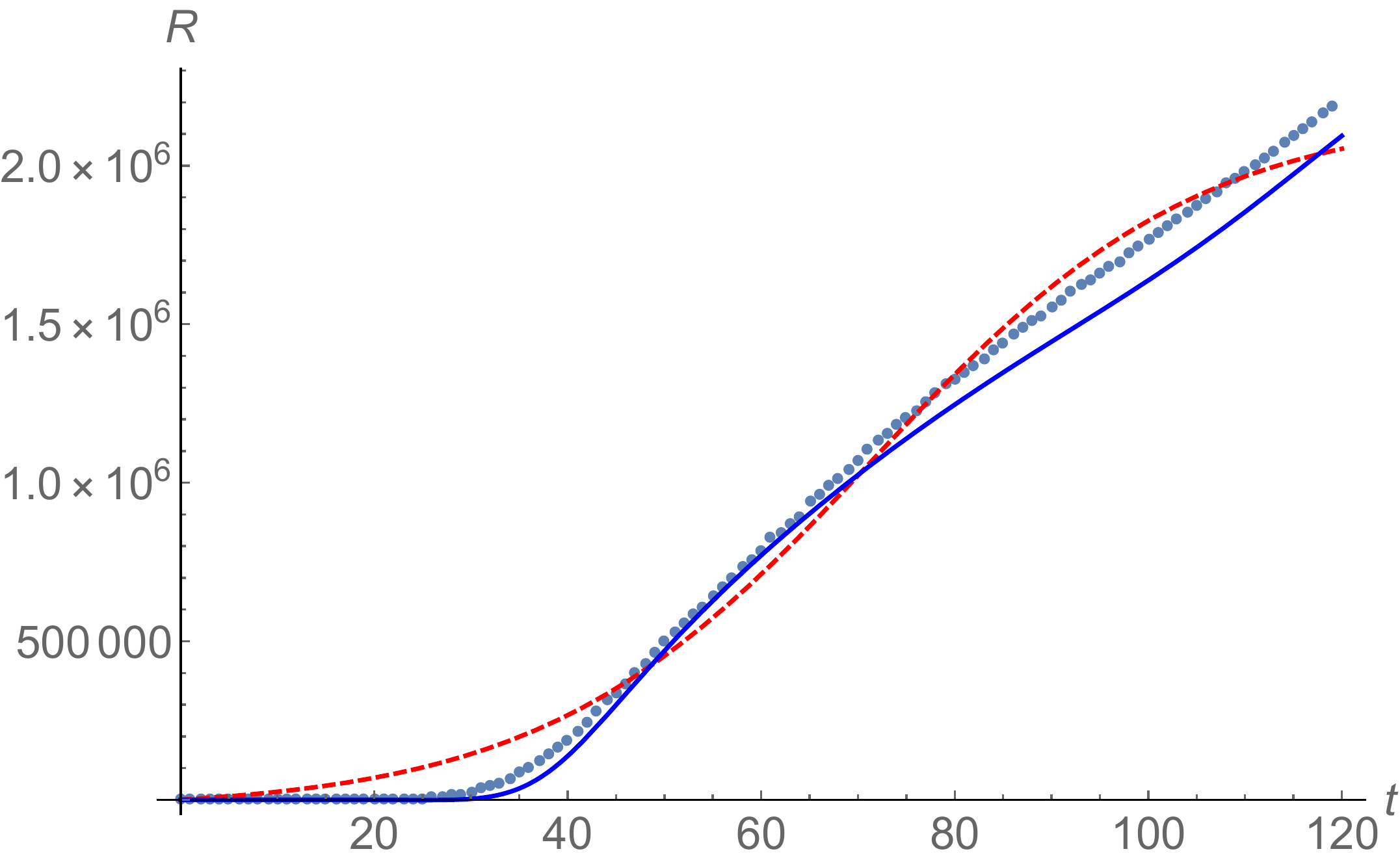} &   \includegraphics[width=80mm]{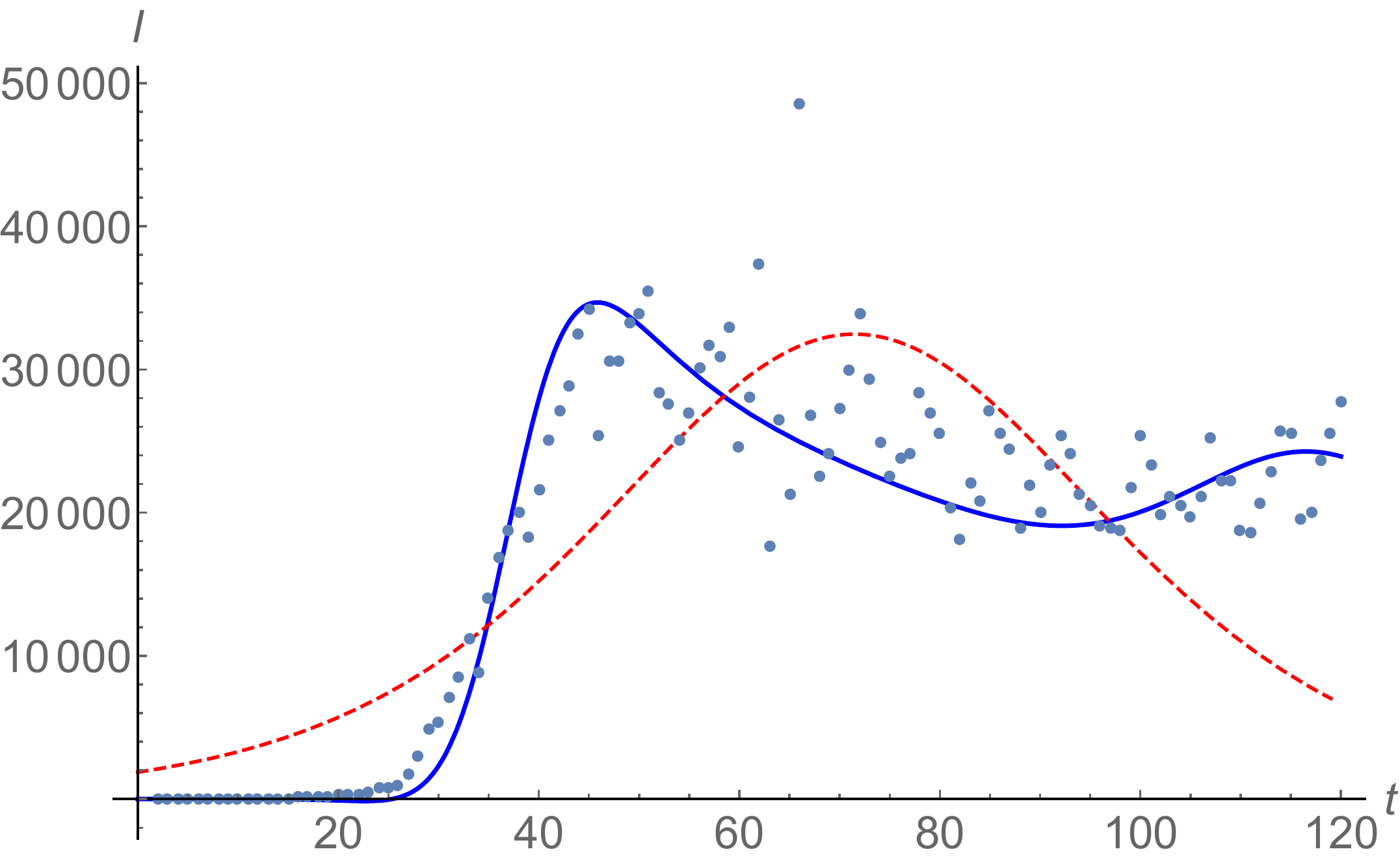} \\
(c) Recovered population for United Sates & (d)  Infected population for United States  \\[6pt]
  \includegraphics[width=80mm]{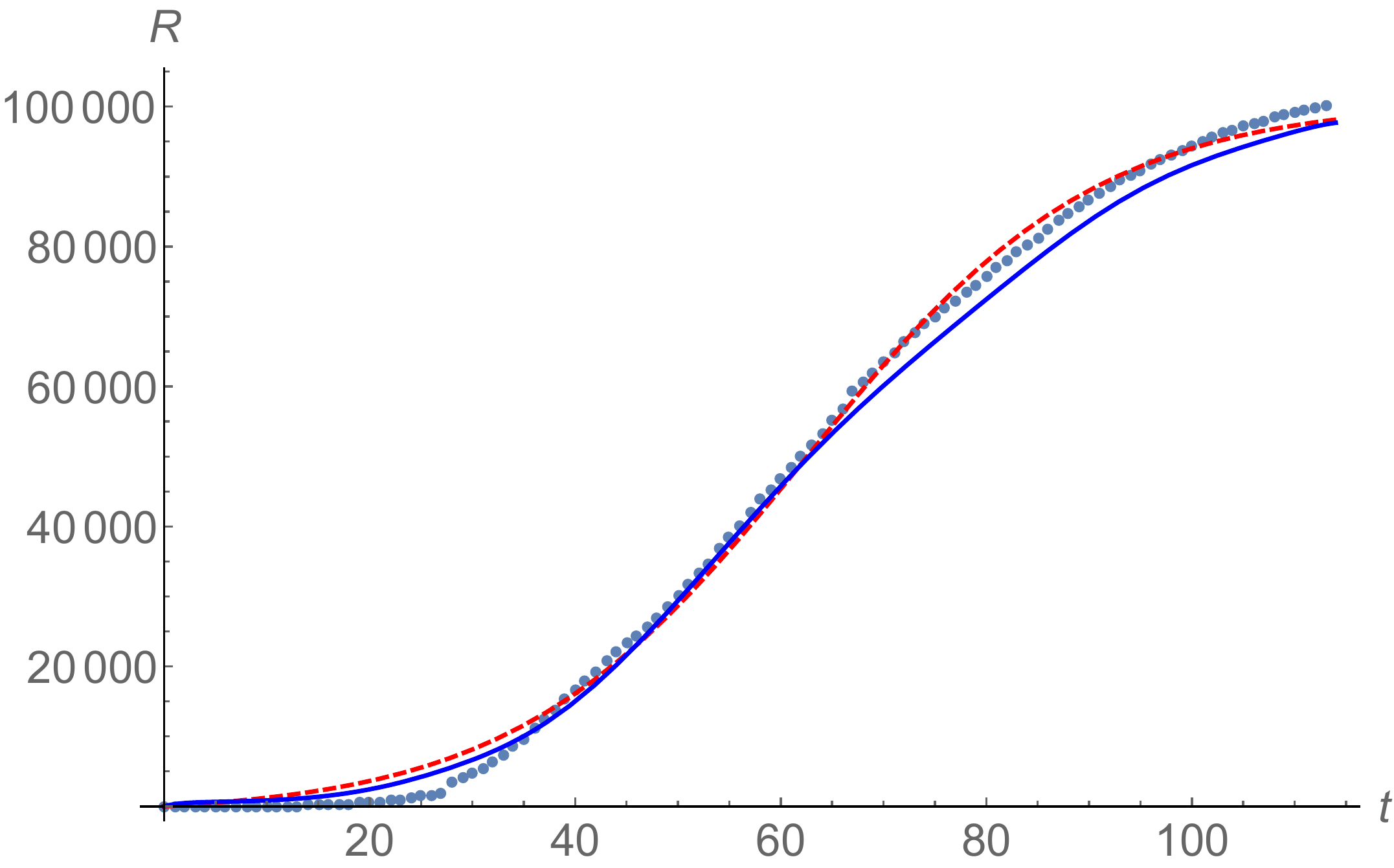} &   \includegraphics[width=80mm]{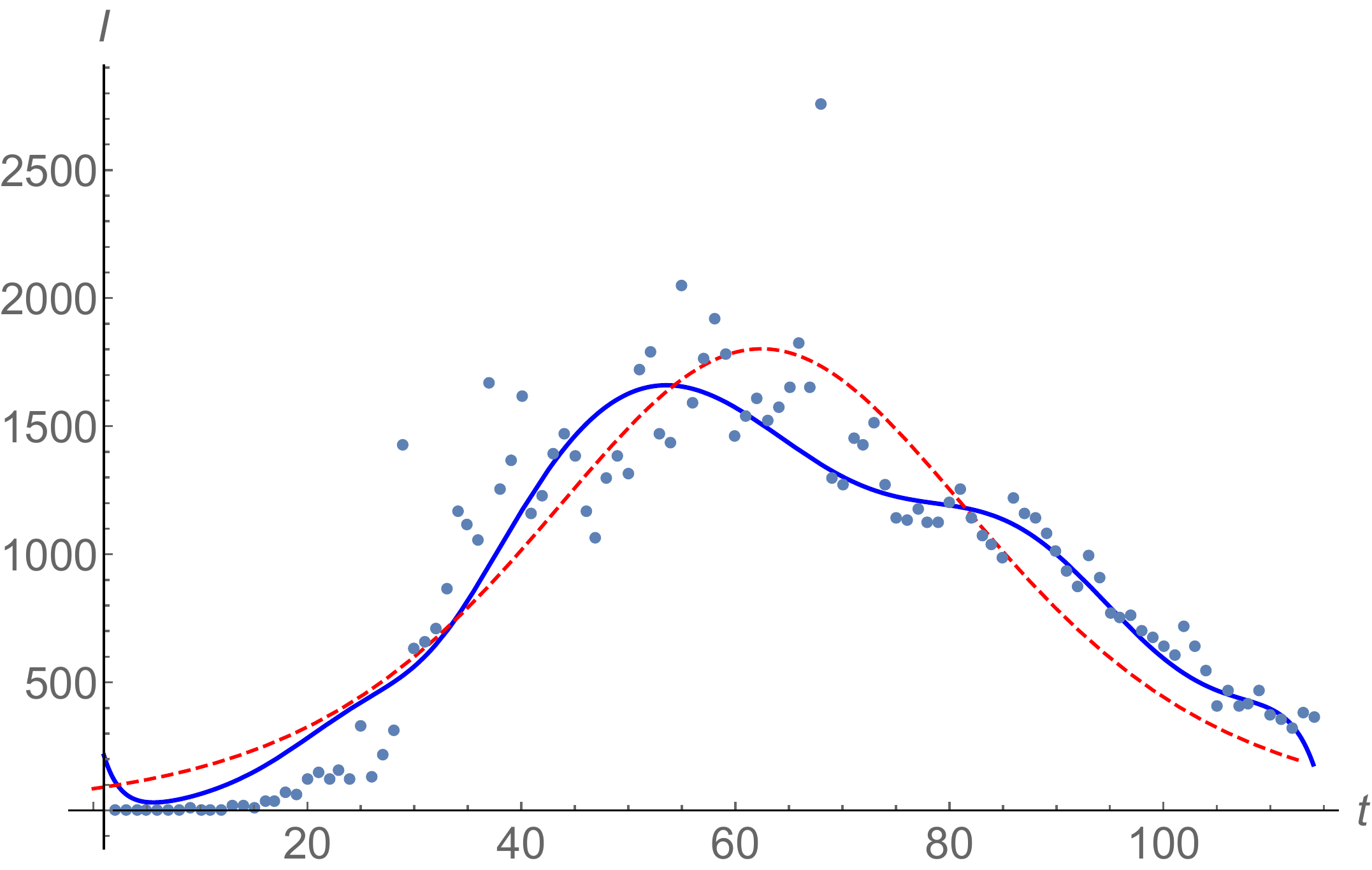} \\
(e) Recovered population for Canada & (f)  Infected population for Canada  \\[6pt]
\includegraphics[width=80mm]{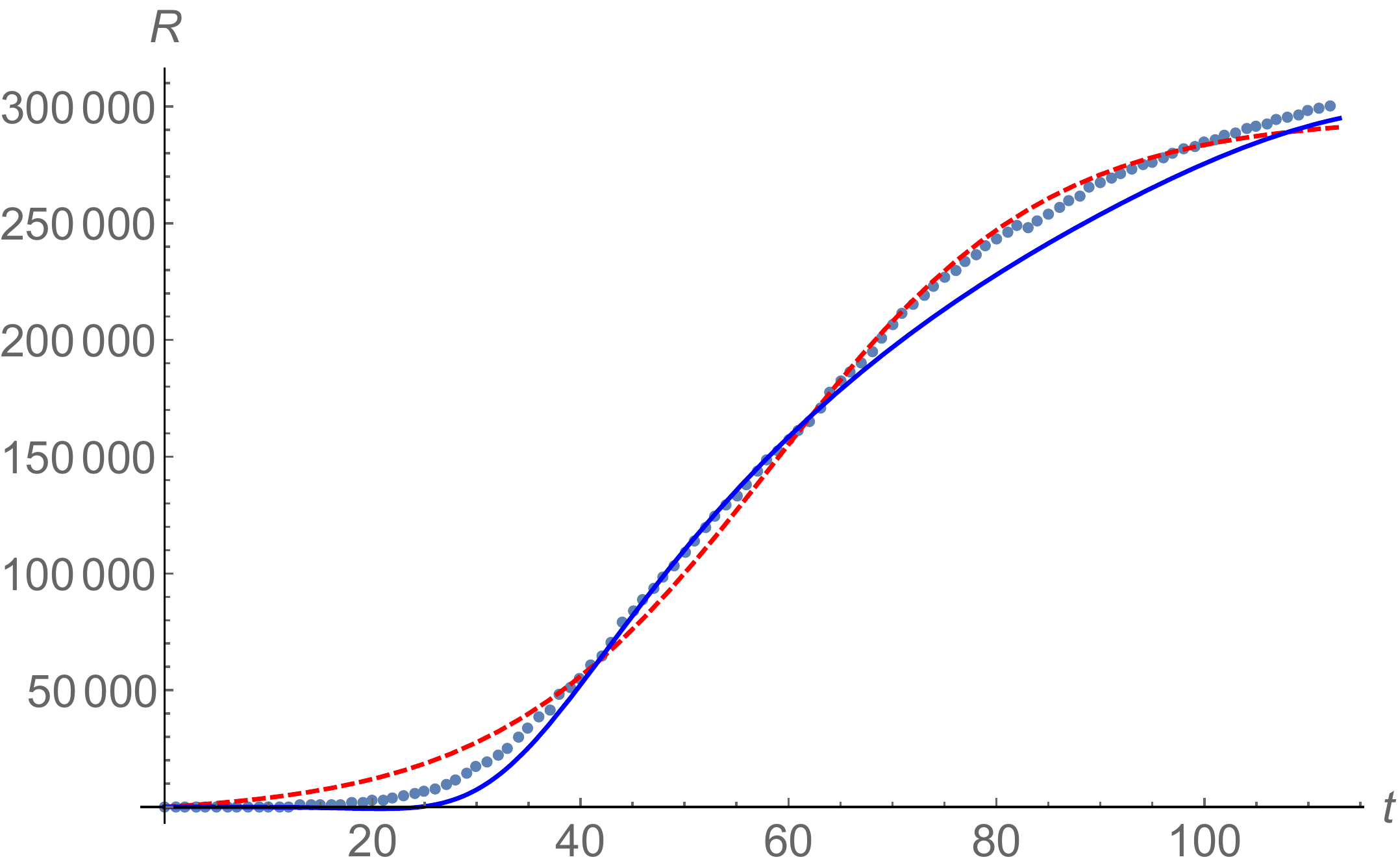} &   \includegraphics[width=80mm]{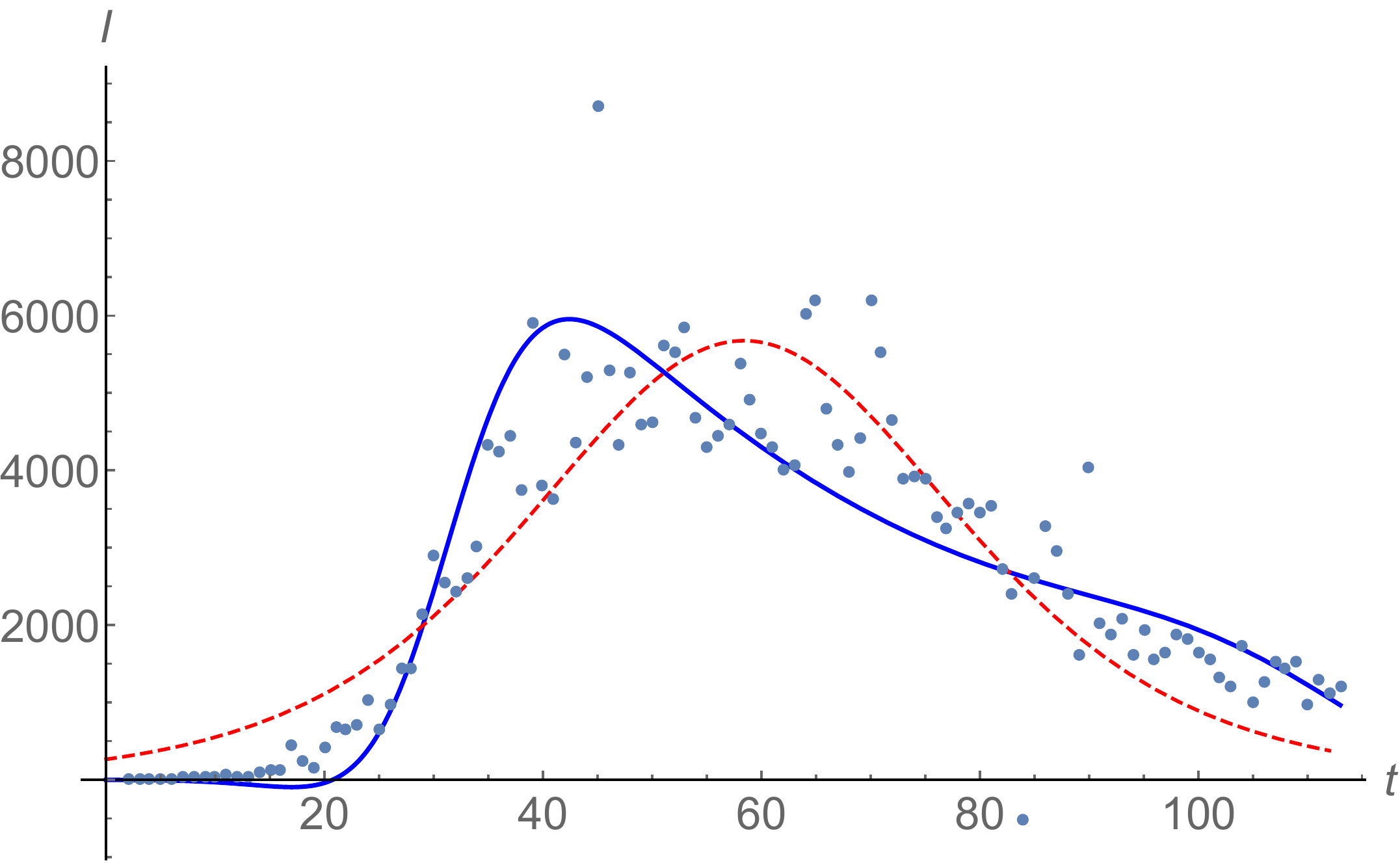} \\
(g) Recovered population for United Kingdom & (h)  Infected population for United Kingdom  \\[6pt]
\end{tabular}
\caption{Recovered and Infected populations for  Italy,  United Sates, Canada and United Kingdom. Data is shown in dotted line.
Fittings \eqref{Req1b} and \eqref{Ieq1} are shown in red dashed line for recovered and and infected populations, respectively. Phenomenological fittings  \eqref{Req2} and \eqref{Ieq2}, with meta--parameters, are shown in blue solid  line for recovered and and infected populations, respectively.}
\label{datasRI}
\end{figure*}

\begin{figure*}[ht]
\begin{tabular}{c}
  \includegraphics[width=160mm]{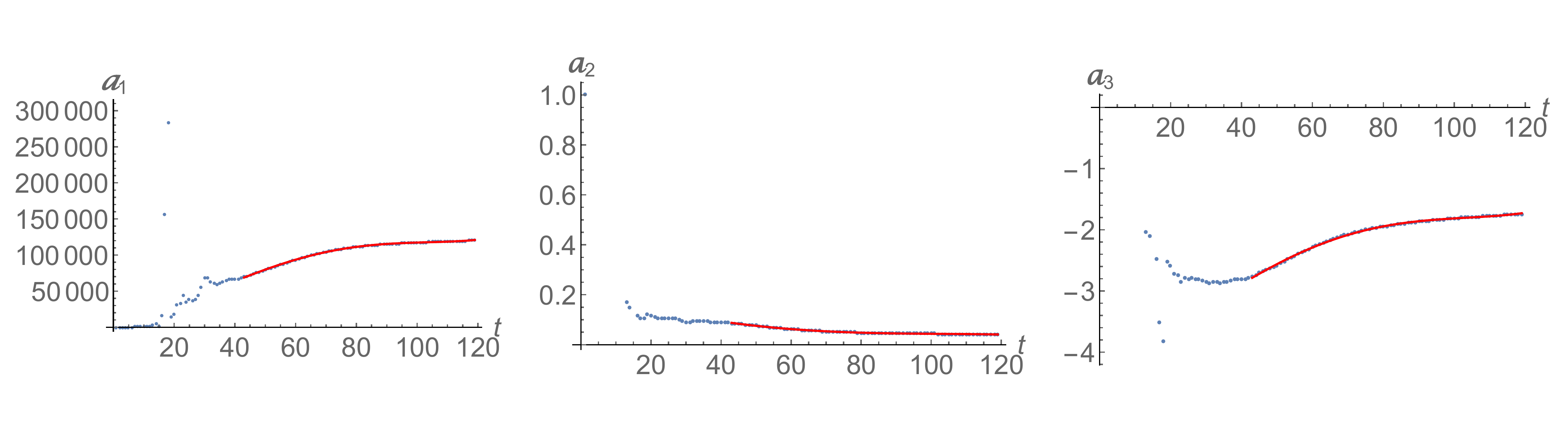} \\
(a) Meta--parameters $a_1$, $a_2$, and $a_3$ for Italy \\[6pt]
  \includegraphics[width=160mm]{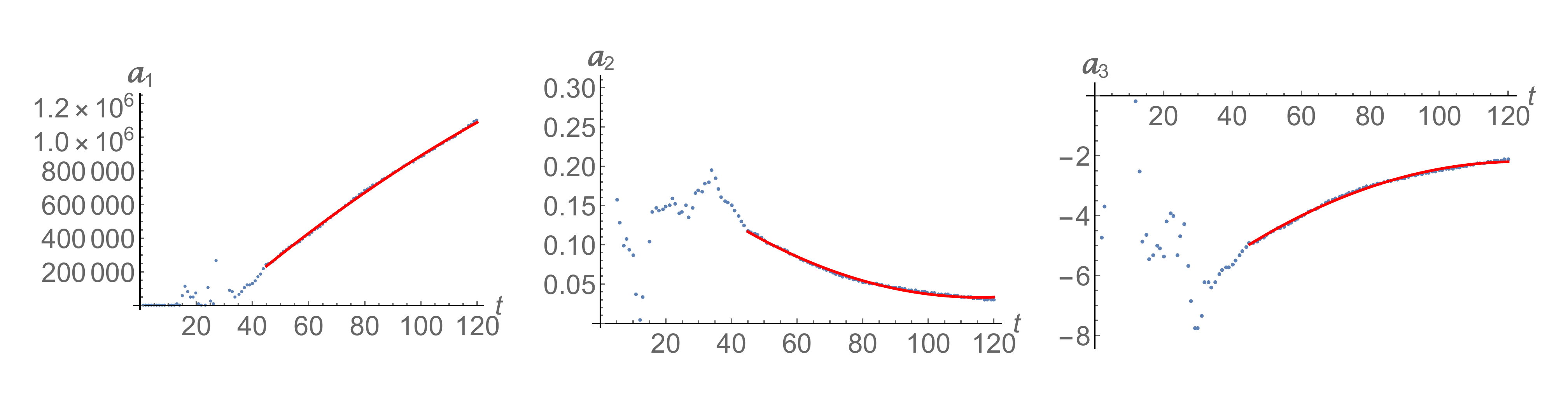} \\
(b) Meta--parameters $a_1$, $a_2$, and $a_3$ for United States \\[6pt]
 \includegraphics[width=160mm]{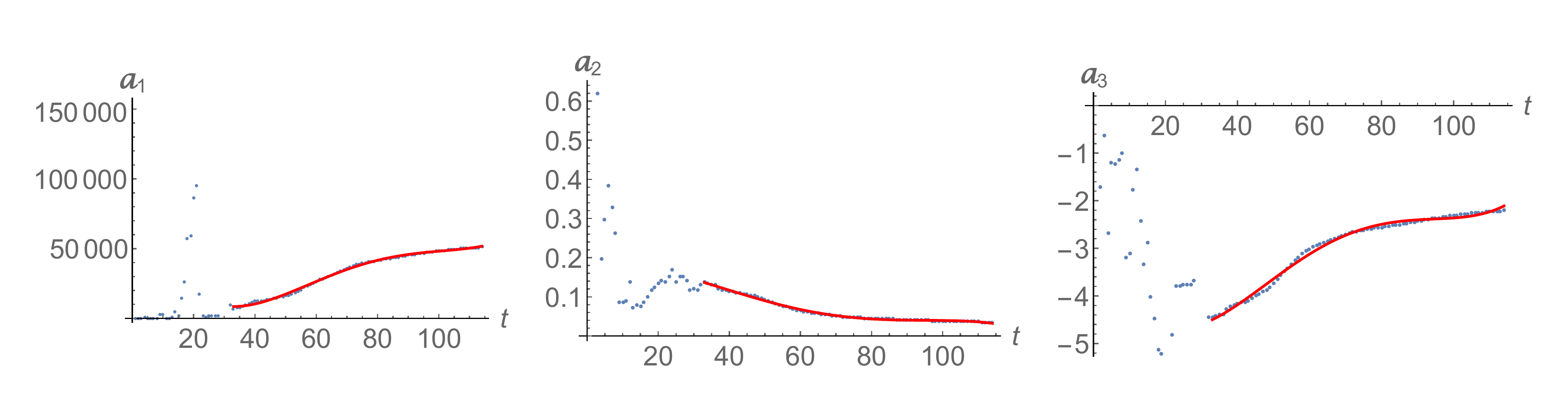} \\
(c) Meta--parameters $a_1$, $a_2$, and $a_3$ for Canada \\[6pt]
\includegraphics[width=160mm]{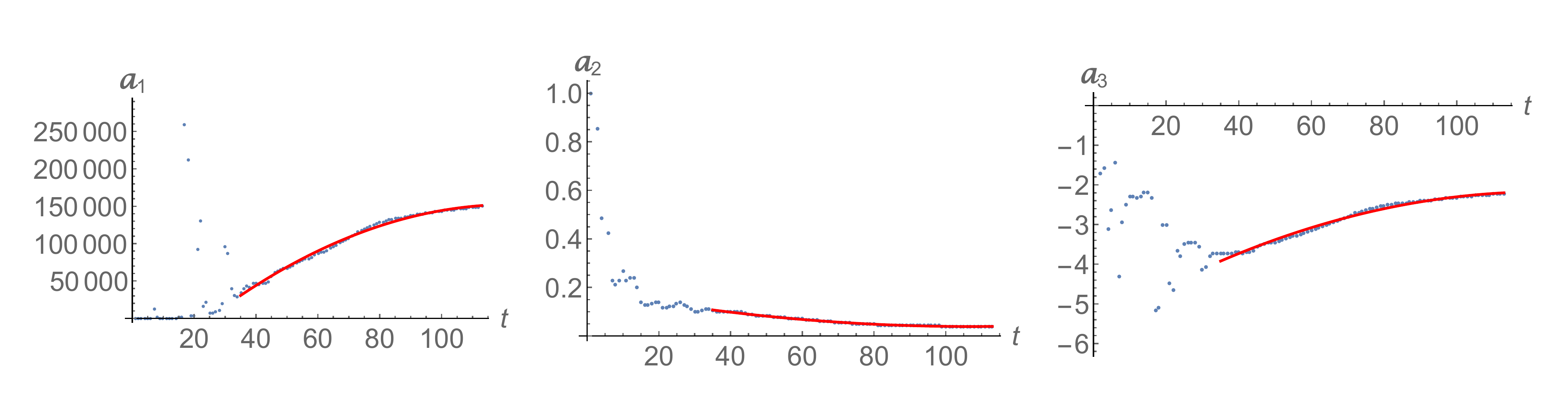} \\
(d) Meta--parameters $a_1$, $a_2$, and $a_3$ for United Kingdom \\[6pt]
\end{tabular}
\caption{Meta--parameters for Italy,  United Sates, Canada and United Kingdom. Data is shown in dotted line.
The regular dynamical behavior of meta--parameters (starting in time $\zeta$) is shown in  red solid  line. They evolve following the quadratic form \eqref{formmetaparametera}.}
\label{datasM1}
\end{figure*}

\subsection{Italy}

The data for the recovered population in Italy is shown by the dotted line in Fig.~\ref{datasRI}(a), with $N=119$. The red-dashed line shows the fitting of function \eqref{Req1b}, with 
parameters $R_0=17$, $a=120072$, $b= 0.04143$ and $c= -1.74422$.

Consider now the blue solid line in Fig.~\ref{datasRI}(a). It describes the meta--analysis fitting \eqref{Req2} for the recovered population since the day $\tau=43$. This is the day in which the meta--parameters start to behave regularly, as it can be  seen in Fig.~\ref{datasM1}(a). The meta--parameters are calculated for each day (from $t=0$ for $R_0$), taking into account all previous days. Thus, each new calculated meta--parameter contains the information of any previous change. Before day $\tau$, there is no regular pattern in the evolution of meta--parameter. But, after day $\tau$, a very distinctive regular dynamical behavior emerge. 
For the current case, after day $\tau=43$, the meta--parameters $a_1(t)$, $a_2(t)$, and $a_3(t)$
time--dependence behavior \eqref{formmetaparametera}  is obtained for $Z=4$, shown by red lines in Fig.~\ref{datasM1}(a). As an example for this case, we display the coefficients of the form \eqref{formmetaparametera}. These are 
$a_{01}= 15076$, $a_{11}=-21.2629$, 
$a_{21}=54.6027$, $a_{31}=0.68966$, $a_{41}=0.00248$,     
$a_{02}= 0.21857$, $a_{12}=-0.00426$, 
$a_{22}= 0.000028$, $a_{32}=2.97537$, $a_{42}=-5.8174\times 10^{-10}$,     
$a_{03}= -4.25959$, $a_{13}=0.02251$, 
$a_{23}= 0.00063$, $a_{33}=-9.8595\times 10^{-6}$, and $a_{43}=3.7844\times 10^{-8}$.
These meta--parameters are used in \eqref{Req2} 
to obtain the solid blue fit in Fig.~\ref{datasRI}(a).

Notice that the fitting precision for the slope of the recovered population becomes more accurate  as it grows, i.e., a better fit for its derivative is achieved, compared to fitting \eqref{Req1b}. This implies that it contains better information on the the number of infected individuals. This is shown in  Fig.~\ref{datasRI}(b), where the data represented by dots is the infected population calculated according to Eq.~\eqref{infecteddataresta}. The red dashed line is the infected population fit \eqref{Ieq1}, with their respectively parameters. Notice also that this fit is just proportional to the approximated solution for infected population in the SIR model.
On the other hand, the solid blue line in Fig.~\ref{datasRI}(b) corresponds to the meta--analysis fitting \eqref{Ieq2} for infected population, with meta--parameters described by the above coefficients.
This fitting reproduces better the global behavior of the evolution of infected population in Italy, just by considering the last part  of the growing on recovered cases (for times $t>\tau$).

The evaluation of the global adjustment function gives
$\epsilon=0.39021$, allowing us to establish that the meta--parameter fitting for infected population is superior to fit \eqref{Ieq1}.

\subsection{United States}

The recovered population for United States is shown in Fig.~\ref{datasRI}(c).  The data with $N=120$ is represented by a dotted line, while the red dashed line shows the fitting  \eqref{Req1b}, 
with parameters $R_0=16$, $a= 243386$, $b= 0.11813$, $c= -4.93365$.
The blue solid line is the fitting \eqref{Req2} since $\tau=45$, when the meta--parameters start to have a regular evolution \eqref{formmetaparametera} with $Z=2$,
 as it can be seen in Fig.~\ref{datasM1}(b) in red line \cite{note}. 

In this case, something similar to the previous case occurs. The fitting \eqref{Req2}, with their respective meta--parameters, is not much better than fit \eqref{Req1b} for the recovered population. However, its slope is in much better agreement with the growing rate for the recovered data. This implies that our meta--analysis gives a better fit
for the infected population, compared to the extracted data from Eq.~\eqref{infecteddataresta}. 
This can be seen in Fig.~\ref{datasRI}(d). In this case, the red dashed line represents  the fitting \eqref{Ieq1} for infected population, while blue solid line is our meta-analysis fitting \eqref{Ieq2} using meta--parameters \eqref{formmetaparametera}.

The fit due to meta--parameters
 is so dramatic, that when the global solution \eqref{Ieq1}  shows a decreasing on infected population, the meta--analysis shows that the rate is not slowed down, but it is increasing. For times $t\gtrapprox  80$, our meta--analysis has been able to extract the information of the  sub--epidemics occuring during the complete time scale of the pandemic. Sub--epidemics are ussually modelled as overlaping of several  epidemic episodes. These are usually described as
multimodal epidemic events satisfying coupled differential equation models \cite{chowell}.
However, in our meta--analysis model, a large sub--epidemic behavior emerge from the study of the evolution of meta--parameters, without invoking any specific model designed to described these kind of epidemics. This is a strength of our model, as no a priori knowledge of the evolution of the pandemic is needed in order to discover the sub--epidemic events in a society. In the case of United States, this sub--epidemic event is happening right now.

Furthermore, the meta--analysis fitting is better than model \eqref{Ieq1}, as the global adjustment function is $\epsilon=0.27888$.

\subsection{Canada}

The recovered population data for Canada  is shown with the dotted line in Fig.~\ref{datasRI}(e). The red dashed line is the fit of  Eq.~\eqref{Req1b} with  $N=114$ and parameters $R_0=12$, $a=-50965.9$, $b= 0.03536$, and $c= -2.20783$.

Anew, the blue solid line is our meta--analysis fitting \eqref{Req2}, using the  meta--parameters that are described by a regular dynamics \eqref{formmetaparametera}, with $Z=4$ and
starting in $\tau=33$, as red lines in Figs.~\ref{datasM1}(c).  
The blue solid line for recovered cases indicates a better approximation to the growing (slope) of such data. 

The infected population is depicted in Fig.~\ref{datasRI}(f), where the data is obtained from \eqref{infecteddataresta}, while
in red dashed line is the fitting \eqref{Ieq1}, and in blue solid line we have the  fitting \eqref{Ieq2}  with  the meta--parameters. 
Once again, the meta--analysis shows several sub--epidemics events, that cannot be obtained by fitting \eqref{Ieq1}. This sub-epidemics appears only when the time--dependent evolution of meta--parameter of Figs.~\ref{datasM1}(c) are found.

Finally, with $\epsilon=0.81282$ for this case, the meta--analysis represents a better fit to the global evolution of infected population.

\subsection{United Kingdom}

The recovered population data is shown in Fig.~\ref{datasRI}(g), while the infected population data is shown in 
Fig.~\ref{datasRI}(h), both of them in dotted lines. The  recovered population data fit 
\eqref{Req1b}, in red line in Fig.~\ref{datasRI}(g), is achieved with $N=113$ and parameters $R_0=16$, $a= 149676$, $b= 0.03791$, and $c =-2.21432$. 

The blue solid line in Fig.~\ref{datasRI}(g), represents the fit of meta--analysis \eqref{Req2} for meta--parameters \eqref{formmetaparametera}
 that have achieved regular evolution for $\tau=35$ and $Z=2$
[see Figs.~\ref{datasM1}(d)].  The information in the meta--analysis is used for infected population in comparison with data \eqref{infecteddataresta}. In   Fig.~\ref{datasRI}(h), the solid blue line for meta--parameters fitting \eqref{Ieq2} shows a better correspondence than fit \eqref{Ieq1} with constant parameters. 
This is confirmed by evaluating $\epsilon=0.84056$.

\subsection{Spain}

In Fig.~\ref{datasRI2}(a) we show the recovered population data (in dotted line). The red dashed line is the fitting \eqref{Req1b} for recovered population with $N=114$, and parameters $R_0=9$, $a=119352$, $b= 0.05597$, and $c=-2.07708$. Similarly, the blue solid line is fitting \eqref{Req2} with meta--parameters \eqref{formmetaparametera} with $Z=4$, and starting in $\tau=35$. We can see in Fig.~\ref{datasM2}(a) that the meta--parameters are described by quartic functions. 

Using this, we can calculate the behavior of the number of the infected individuals, shown in Fig.~\ref{datasRI2}(b) . The infected population data (dotted line) is given by Eq.~\eqref{infecteddataresta}, while fit \eqref{Ieq1} is in red dashed line and meta--analysis fit \eqref{Ieq2} is in solid blue line. We obtain that $\epsilon=0.49292$, showing that the  fit based in meta--parameters is again superior to the one based in time--independent parameters.

\subsection{Poland}

The data for the recovered population of Poland is plotted in 
Fig.~\ref{datasRI2}(c) in dotted line.  Also, the red dashed line  fits Eq.~\eqref{Req1b} for the recovered population with $N=105$, and  parameters   
$R_0=5$, $a=20169.1$, $b=0.02011$, and $c= -1.39525$.
The meta--analysis fitting \eqref{Req2} is in blue line with the meta--parameters \eqref{formmetaparametera} from Fig.~\ref{datasM2}(b),
starting in $\tau=35$, with $Z=2$. 

This case is interesting as the slope of meta--analysis fit \eqref{Req2} clearly shows that recovered population is increasing, as opposed to 
what can be deduced from fit \eqref{Req1b}. Similarly, the infected population data, in Fig.~\ref{datasRI2}(d), shows a better agreement with the fit \eqref{Ieq2}
due to meta--parameters. While the fit \eqref{Ieq1} indicates a strong decreasing on infected population, the meta--analysis shows that infected population is increasing due to sub--epidemic events. 
A better fit of the meta--analysis is also corroborated by $\epsilon=0.65004$.

\subsection{Austria}

\begin{figure*}[ht]
\begin{tabular}{c}
  \includegraphics[width=160mm]{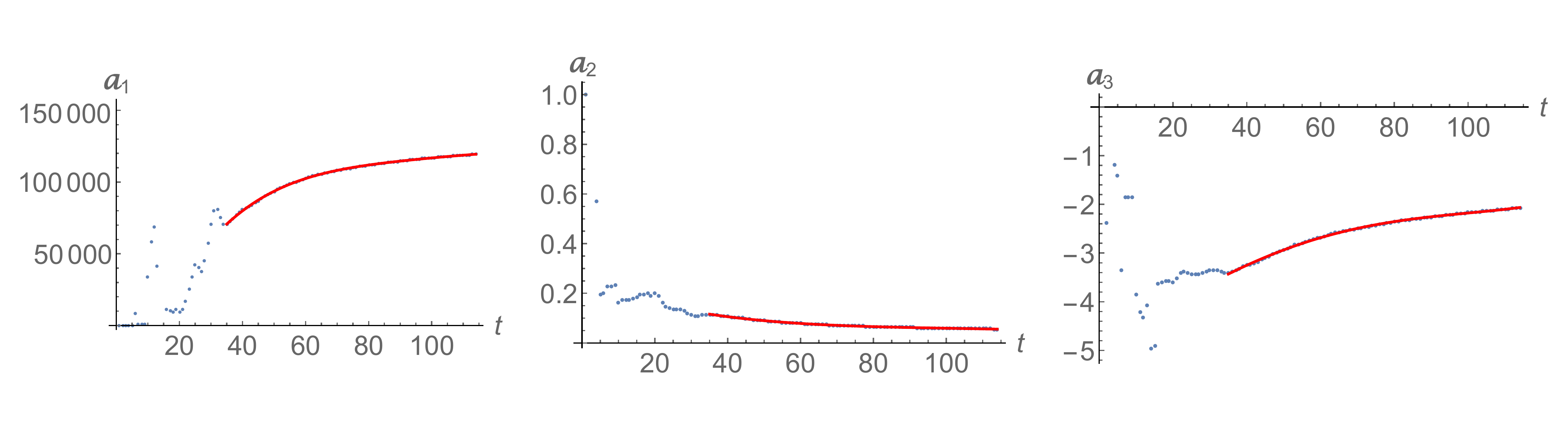} \\
(a) Meta--parameters $a_1$, $a_2$, and $a_3$ for Spain \\[6pt]
 \includegraphics[width=160mm]{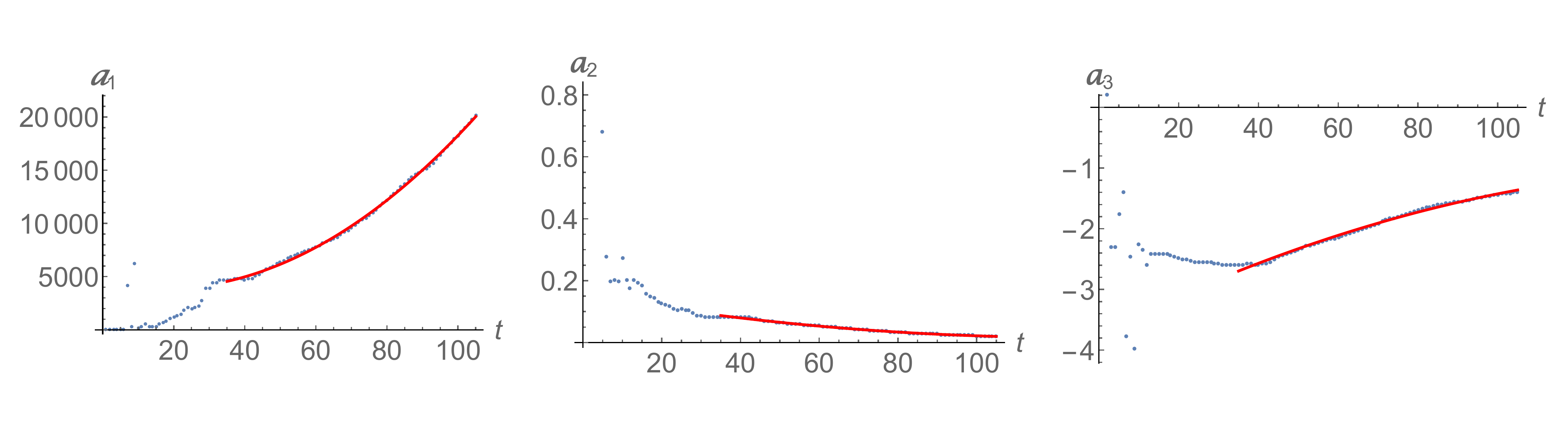} \\
(b) Meta--parameters $a_1$, $a_2$, and $a_3$ for Poland \\[6pt]
\includegraphics[width=160mm]{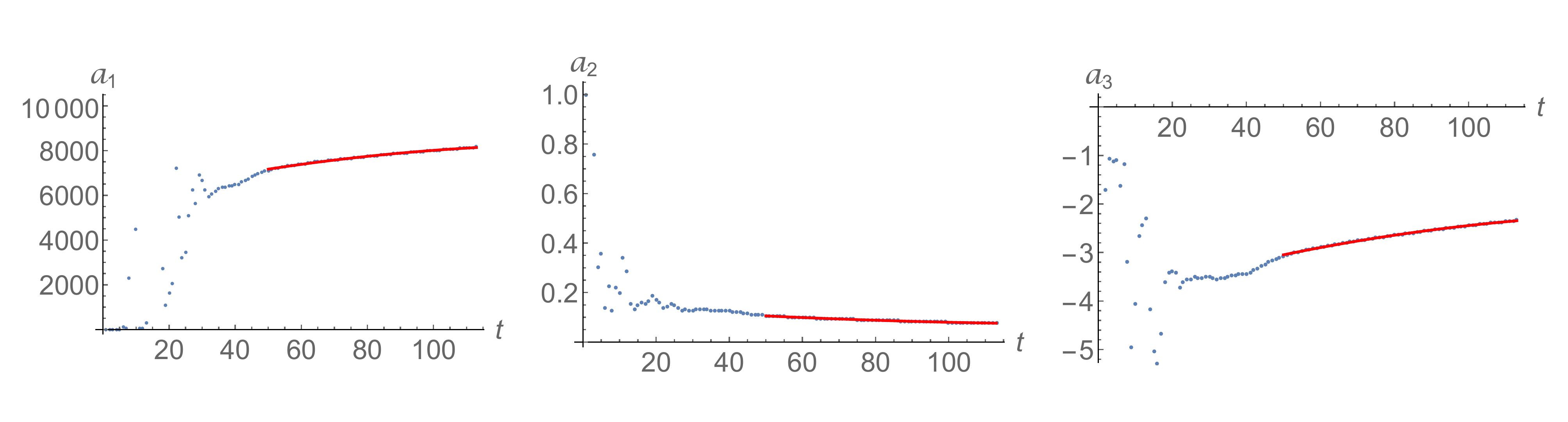} \\
(c) Meta--parameters $a_1$, $a_2$, and $a_3$ for Austria \\[6pt]
\includegraphics[width=160mm]{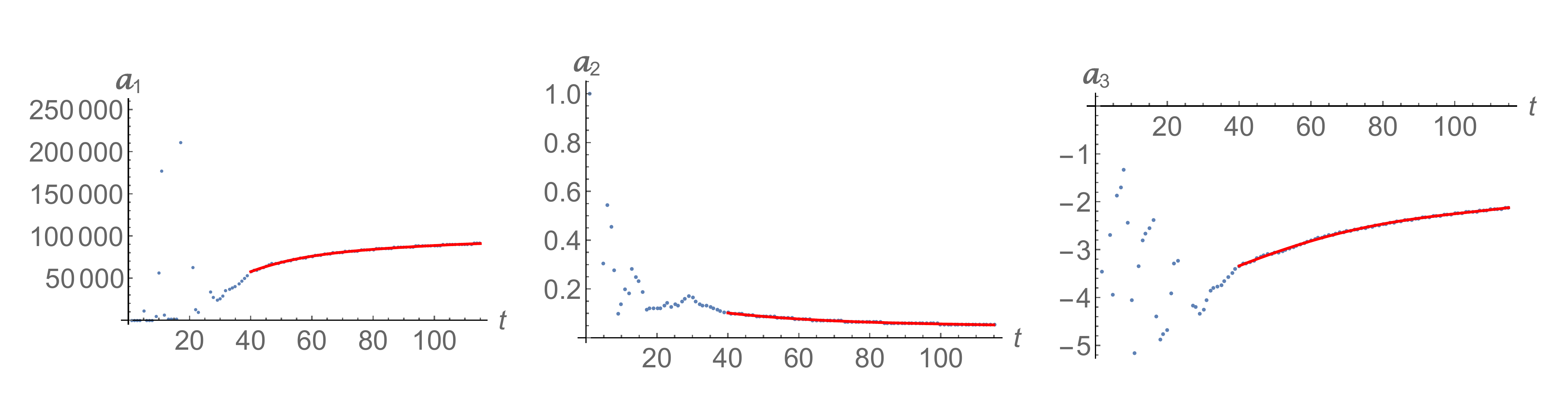} \\
(d) Meta--parameters $a_1$, $a_2$, and $a_3$ for Germany \\[6pt]
\end{tabular}
\caption{Meta--parameters for Spain, Poland, Austria, and Germany. Data is shown in dotted line.
The regular dynamical behavior of each meta--parameter (starting in time $\zeta$) is shown in  red solid  line. They evolve following the quadratic form \eqref{formmetaparametera}.}
\label{datasM2}
\end{figure*}

In Fig.~\ref{datasRI2}(e) is the data for the recovered population in Austria. In red dashed line is the fit \eqref{Req1b} for $N=113$ and  parameters $R_0=5$, $a= 8152.2$, $b= 0.07569$, $c= -2.33693$.
Similarly, the fit \eqref{Req2}, in blue solid line, requires the meta--parameters \eqref{formmetaparametera} show in Fig.~\ref{datasM2}(c),
 with coefficients \eqref{formmetaparametera}, starting from $\tau=50$, with $Z=2$.

The meta--analysis fit indicates a better adjustment for the infected population, as it is seen in Fig.~\ref{datasRI2}(f). In that figure, the red dashed line is the fit \eqref{Ieq1}, and the blue solid line is the fit \eqref{Ieq2}. Notice that better adjustment of the meta--analysis for the tail of the pandemic data, for $t>60$.
The global dynamical behavior of the infected population is better achieved by the meta--analysis, with $\epsilon=0.51678$

\subsection{Germany}

\begin{figure*}[ht]
\begin{tabular}{cc}
  \includegraphics[width=80mm]{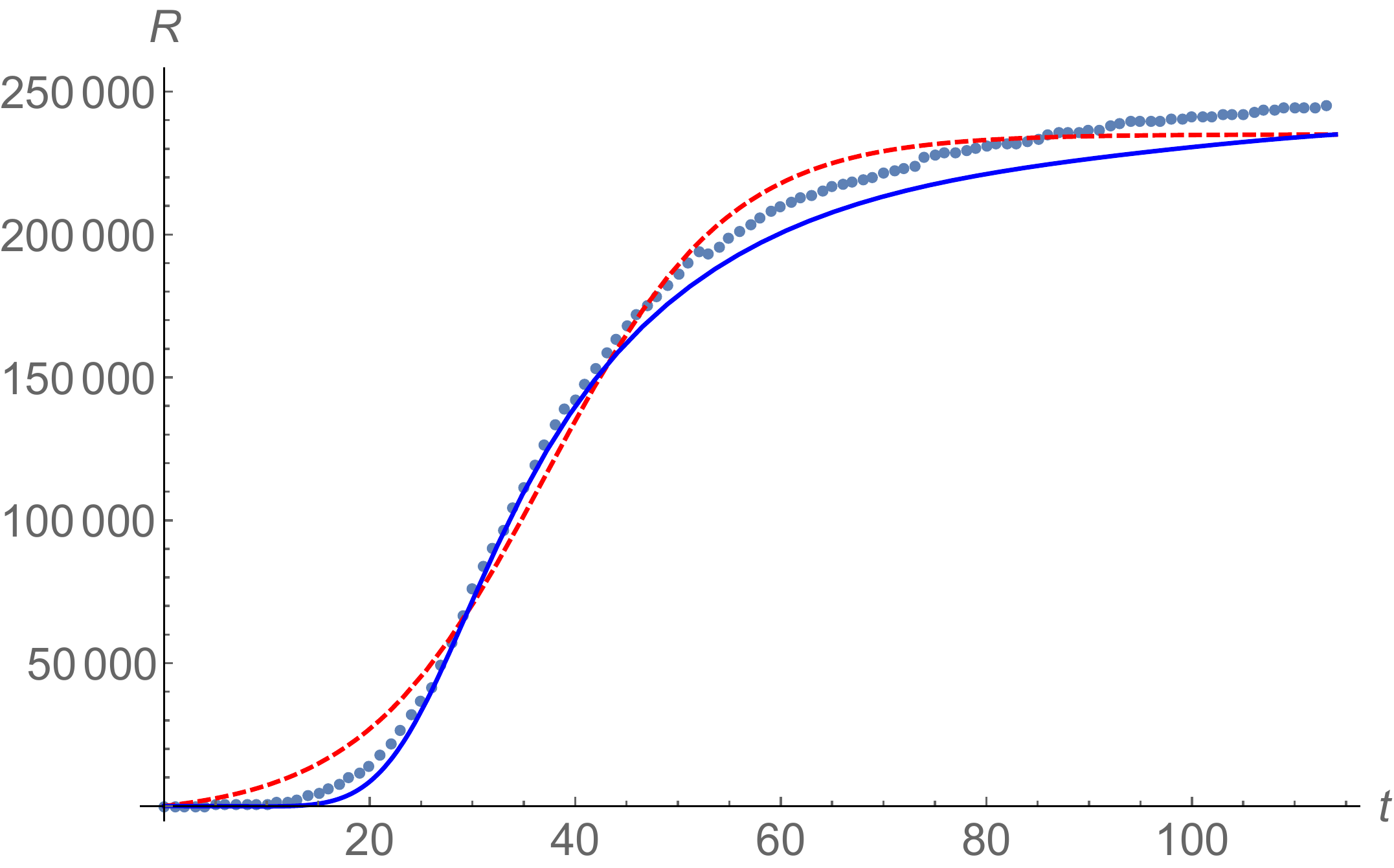} &   \includegraphics[width=80mm]{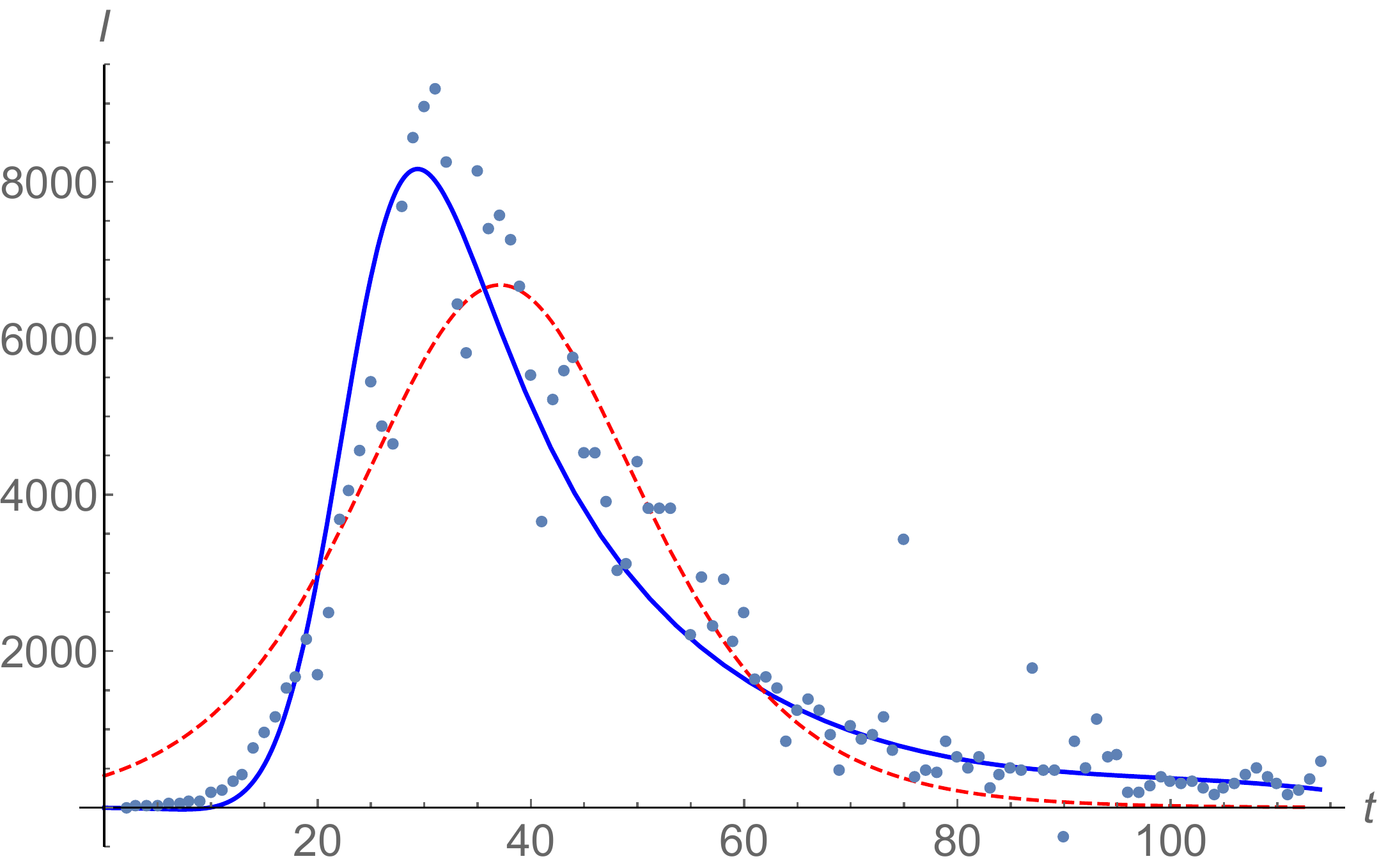} \\
(a) Recovered population for Spain & (b)  Infected population for Spain \\[6pt]
 \includegraphics[width=80mm]{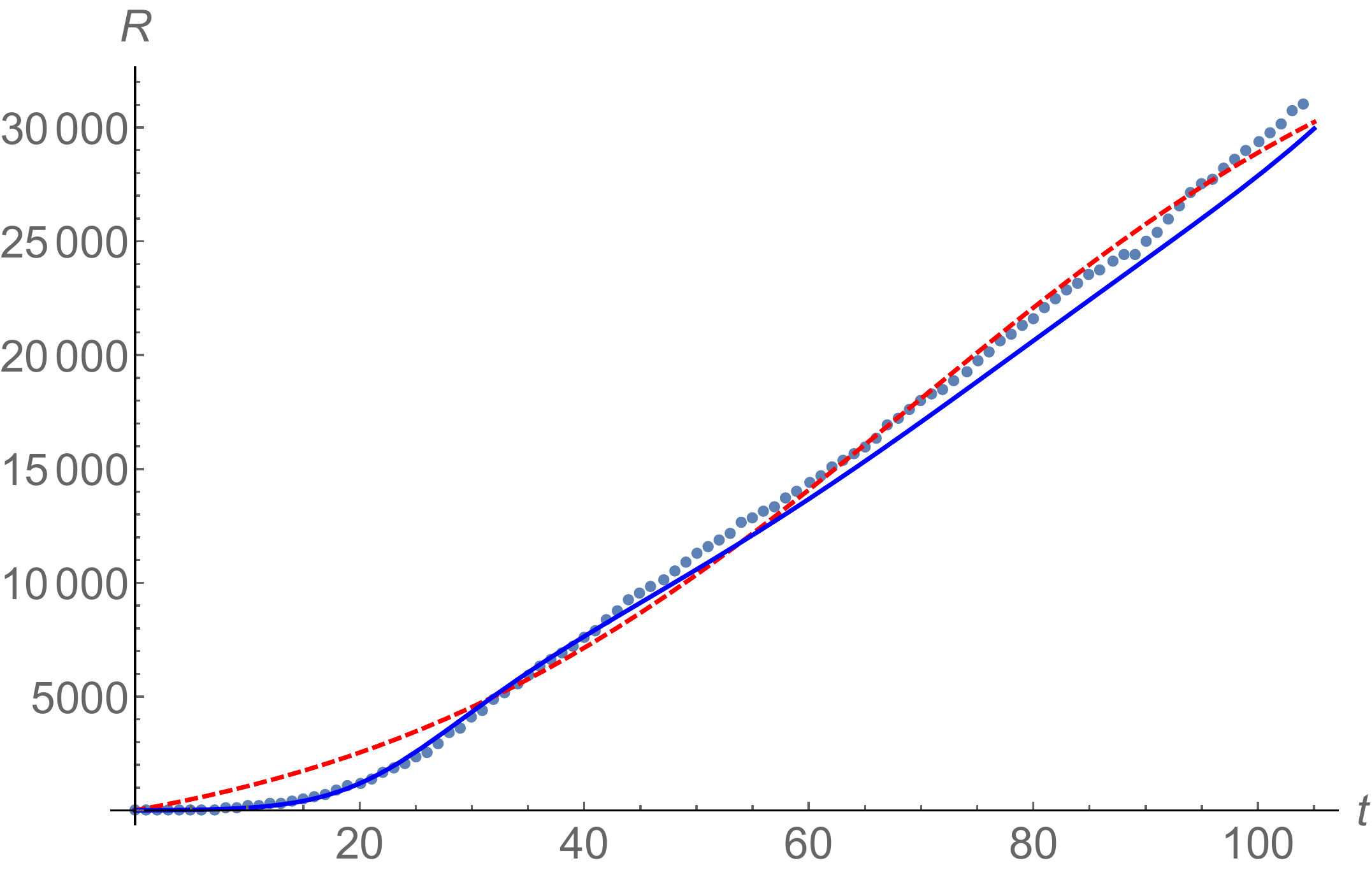} &   \includegraphics[width=80mm]{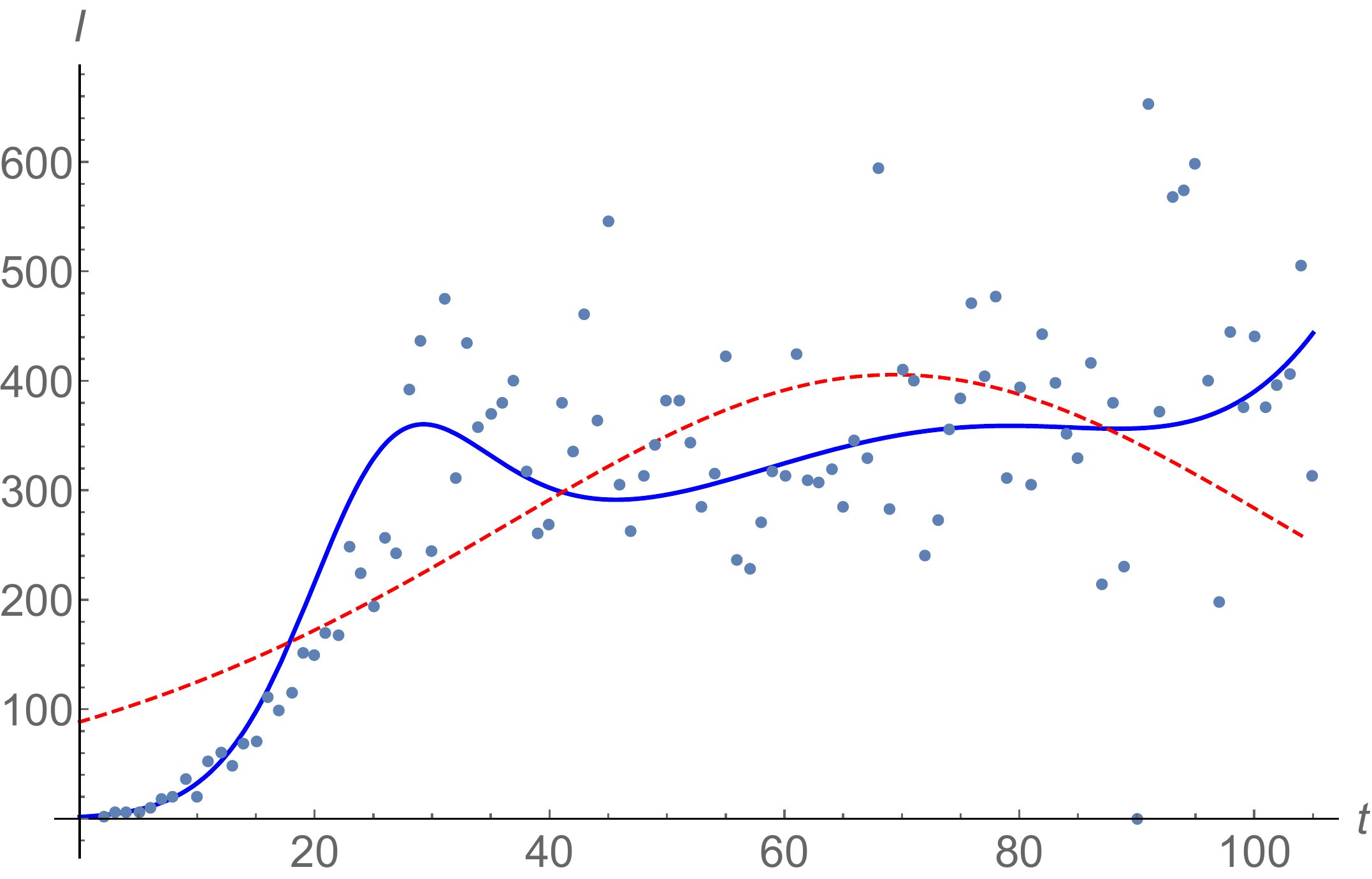} \\
(c) Recovered population for Poland & (d)  Infected population for Poland \\[6pt]
\includegraphics[width=80mm]{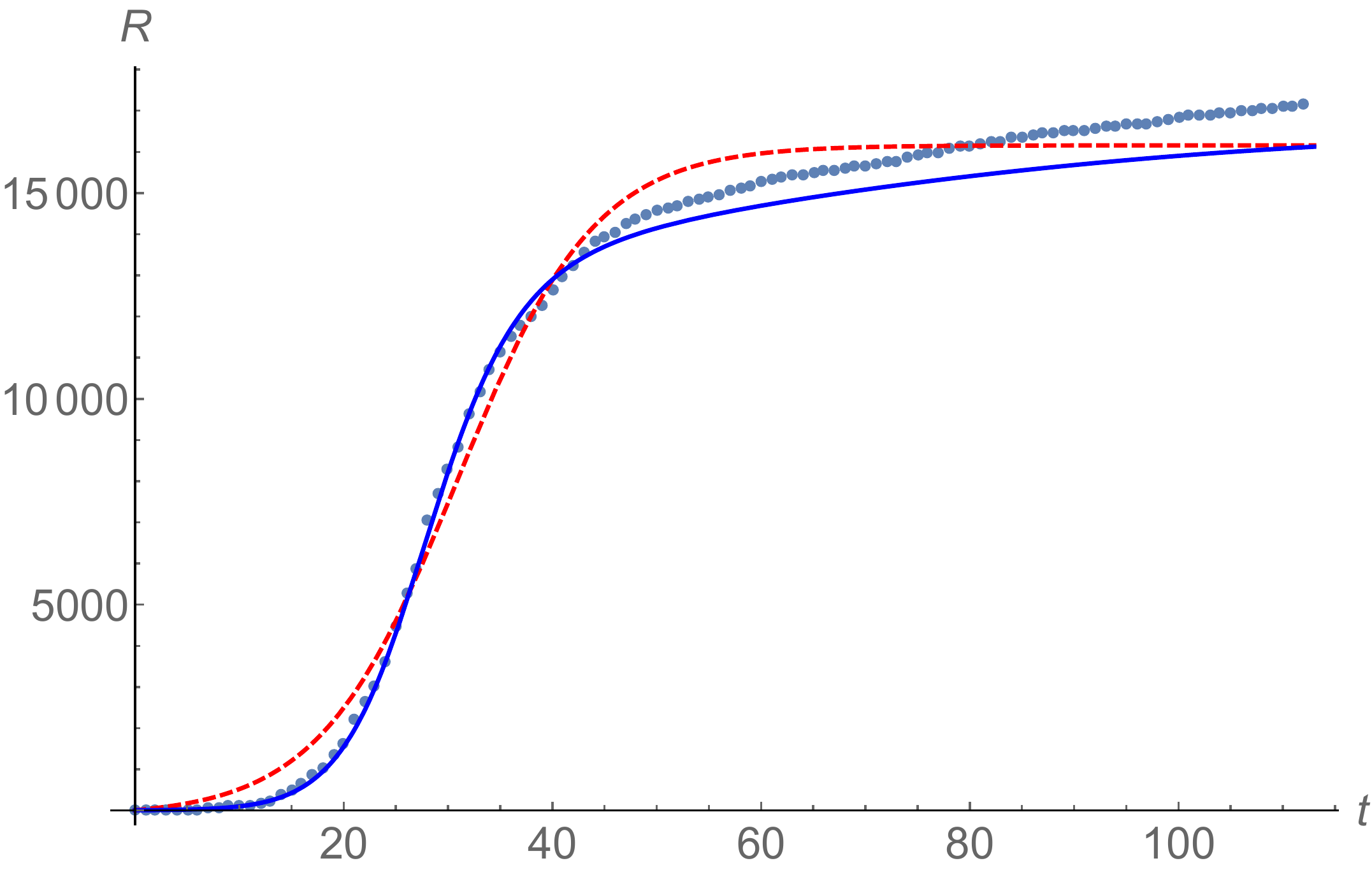} &   \includegraphics[width=80mm]{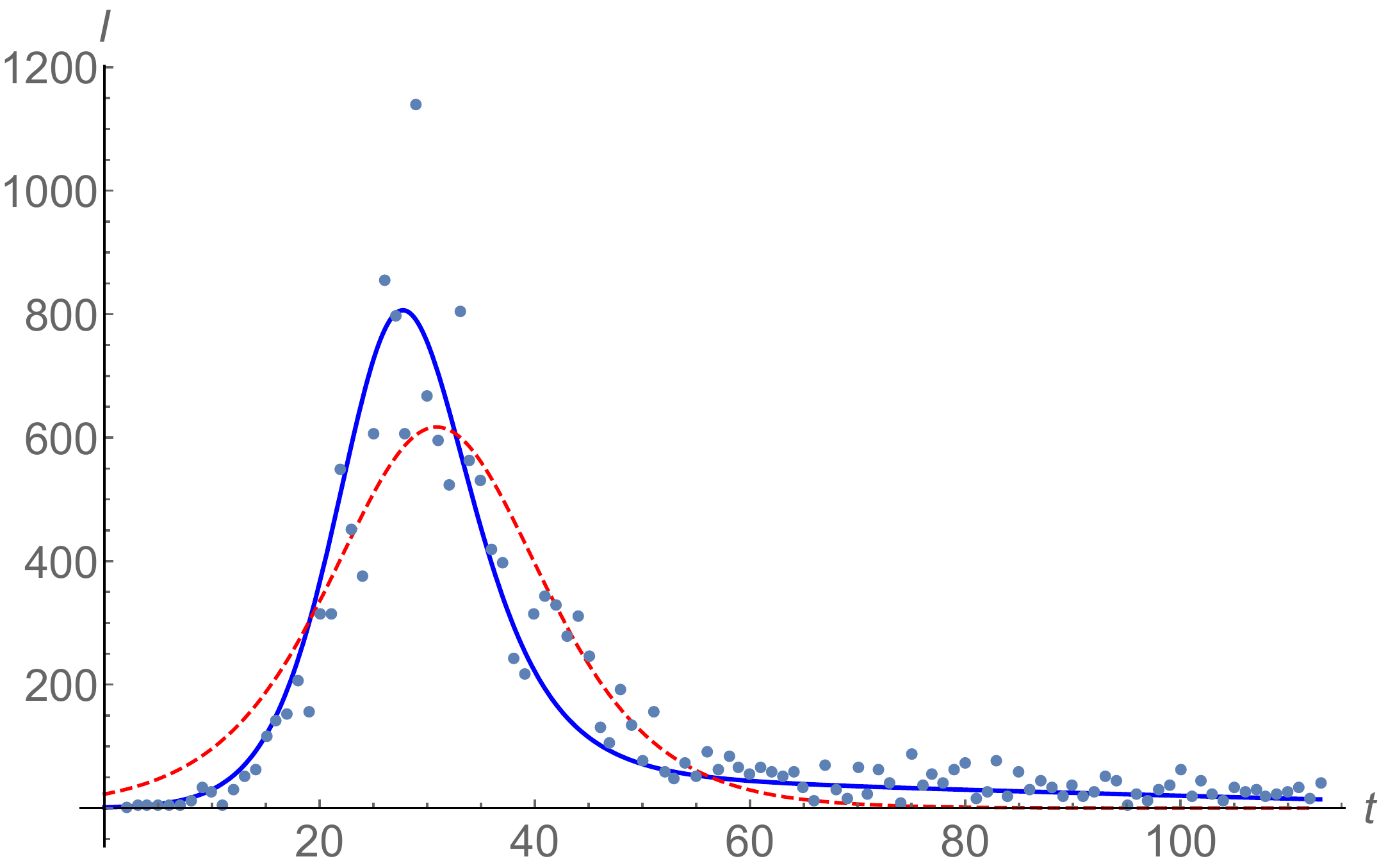} \\
(e) Recovered population for Austria & (f)  Infected population for Austria \\[6pt]
\includegraphics[width=80mm]{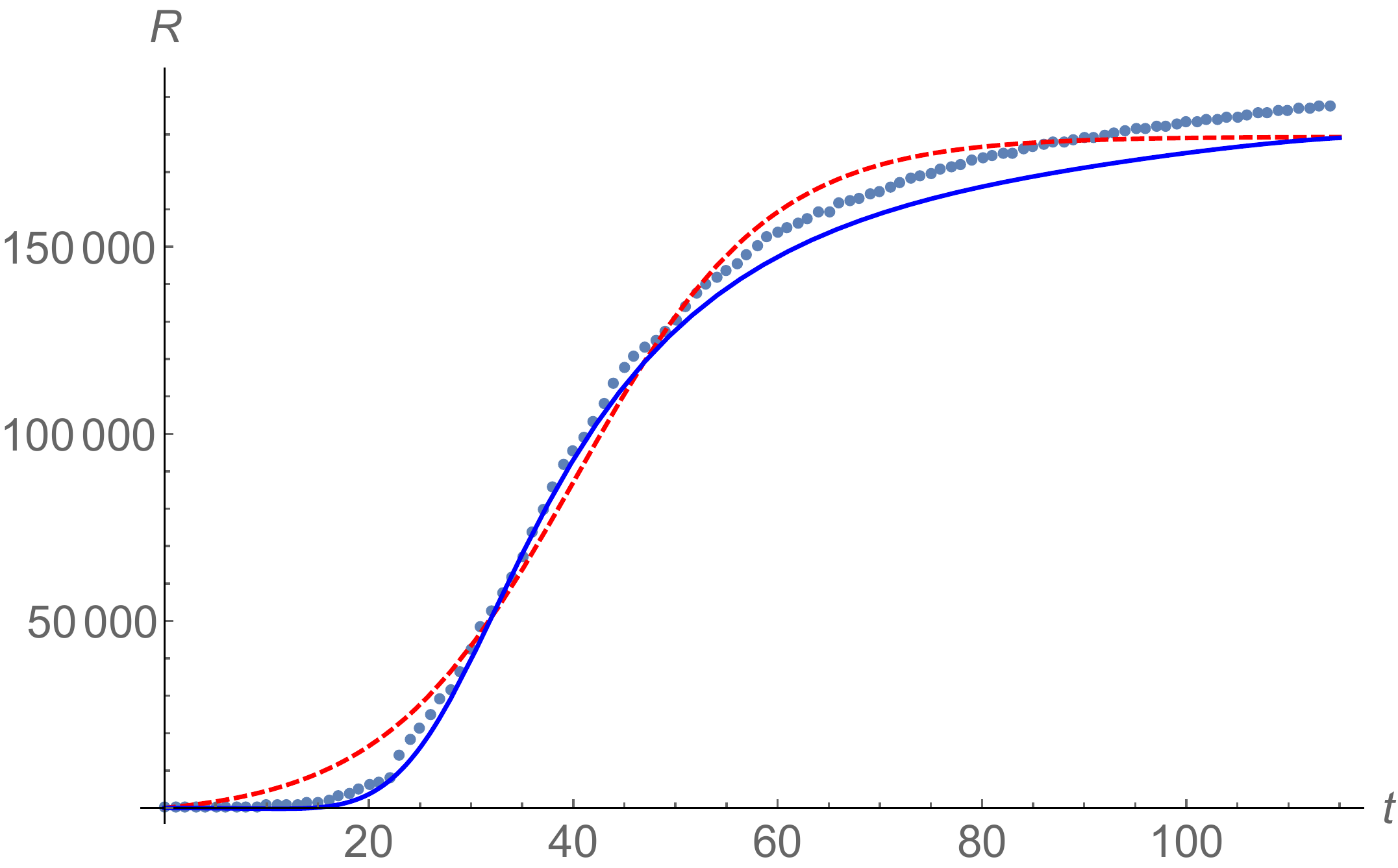} &   \includegraphics[width=80mm]{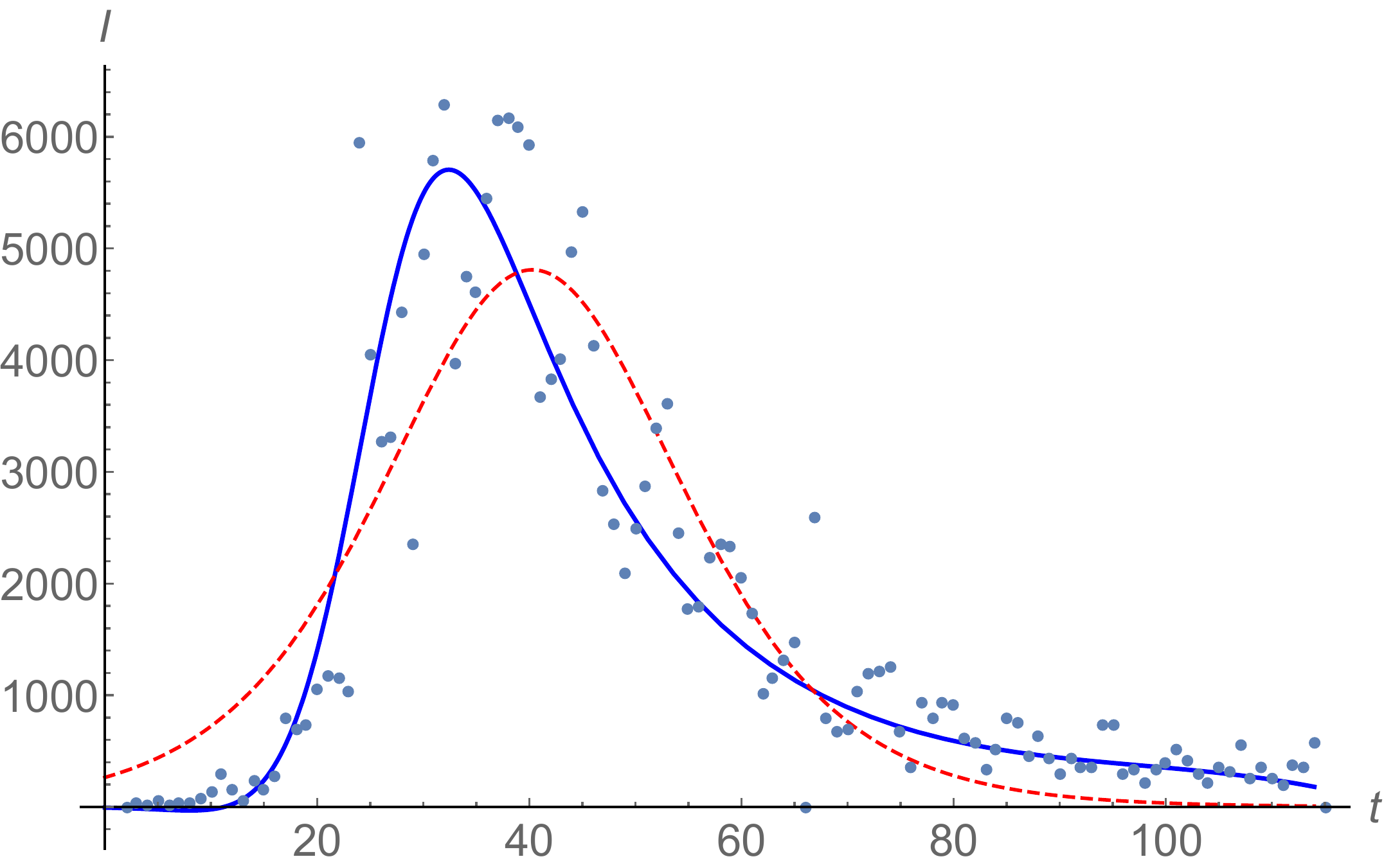} \\
(g) Recovered population for Germany & (h)  Infected population for Germany \\[6pt]
\end{tabular}
\caption{Recovered and Infected populations for Spain, Poland, Austria, and Germany. Data is shown in dotted line.
Fittings \eqref{Req1b} and \eqref{Ieq1} are shown in red dashed line for recovered and and infected populations, respectively. Phenomenological fittings  \eqref{Req2} and \eqref{Ieq2}, with meta--parameters, are shown in blue solid  line for recovered and and infected populations, respectively.}
\label{datasRI2}
\end{figure*}

The data for this country is shown in Figs.~\ref{datasRI2}(g)  and \ref{datasRI2}(h)  for the data (dotted lines) of recovered and infected populations, respectively. Red dashed lines represent the fitting of \eqref{Req1b} and \eqref{Ieq1} for both cases, with $N=115$ and $R_0=17$, $a= 90954.5$, $b= 0.05287$, $c= -2.12998$. Blue solid lines represent the meta--parameter fittings of recovered \eqref{Req2} and infected \eqref{Ieq2} populations, with respect to the data \eqref{infecteddataresta}. The meta--parameters 
 \eqref{formmetaparametera}  start from $\zeta=40$, with $Z=4$.

Once more, the fit \eqref{Ieq2} is better for infected population,  as $\epsilon=0.59191$.

\subsection{Portugal}

In Fig.~\ref{datasRI3}(a) is shown the recovered cases  in dotted line, and the fitting \eqref{Req1b} in red dashed line with $N=106$ and $R_0=9$, $a= 19109.7$, $b= 0.03079$, and $c= -1.25453$. Blue solid line is fit \eqref{Req2}, with metaparamter shown in Figs.~\ref{datasM3}(a), with $\tau=26$ and $Z=4$. Similarly, we show the different fitting  \eqref{Ieq1} and \eqref{Ieq2} for infected populations in  Fig.~\ref{datasRI3}(b). The meta--analysis produces a better fit, as $\epsilon=0.50303$.

Notice that,  again, sub--epidemics events are present, and they are apparent only through the known evolution of meta--parameters.

\subsection{New Zealand}

In Figs.~\ref{datasRI3}(c) and (d) we show the recovered and infected population for New Zealand. In dotted lines are the data, while the fitting \eqref{Req1b} and \eqref{Ieq1} are in red dashed line with $N=94$ and $R_0=20$, $a=606.56$, $b=0.10504$, and $c=-1.28409$.
The blue solid line is the meta--analysis fitting \eqref{Req2} and \eqref{Ieq2}, for meta--paramters of Figs.~\ref{datasM3}(b), with $Z=2$ and $\tau=30$.

This case is very interesting as it is the only one in which the meta--analysis and the fitting with constant parameters almost coincide (although the meta--analysis is better with $\epsilon=0.88445$). This is an indication of the very good health policies adopted by the goverment  to deal with  the pandemic and of the responsible behavior of the society. In this (almost ideal) case, the evolution of the infection variables behaves as a very precise mathematical model.

\subsection{France}

\begin{figure*}[ht]
\begin{tabular}{cc}
  \includegraphics[width=80mm]{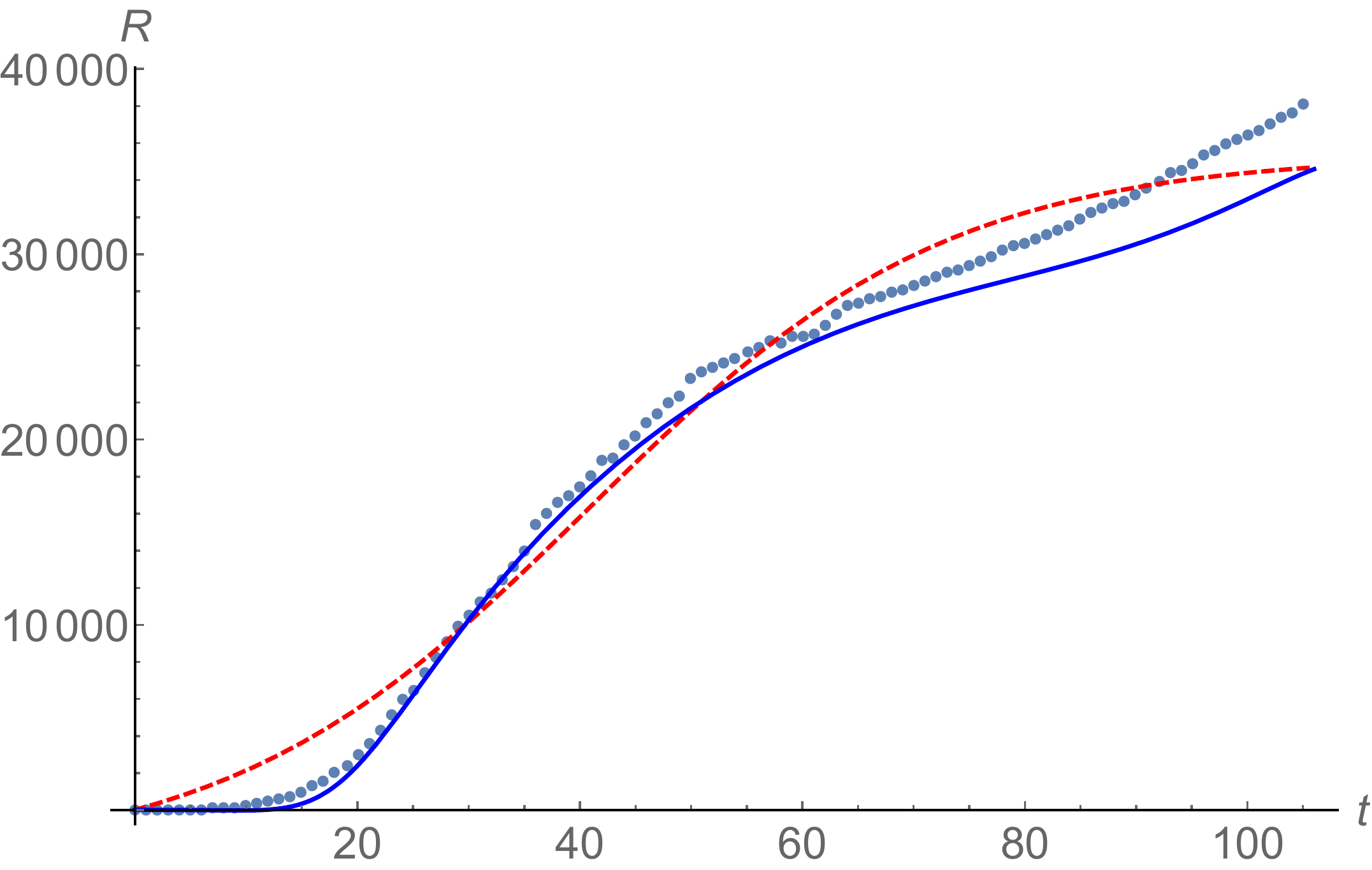} &   \includegraphics[width=80mm]{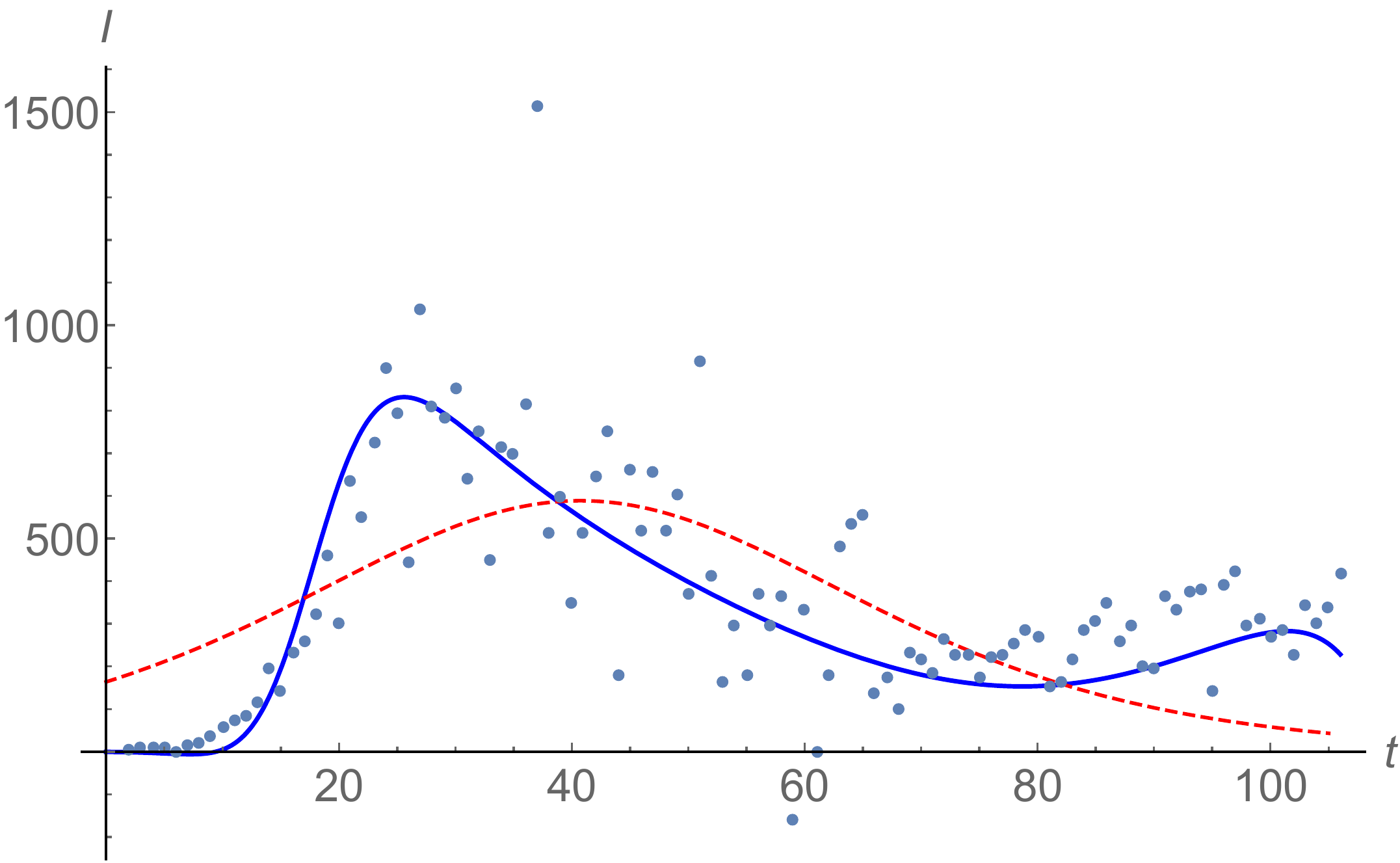} \\
(a) Recovered population for Portugal & (b)  Infected population for Portugal \\[6pt]
 \includegraphics[width=80mm]{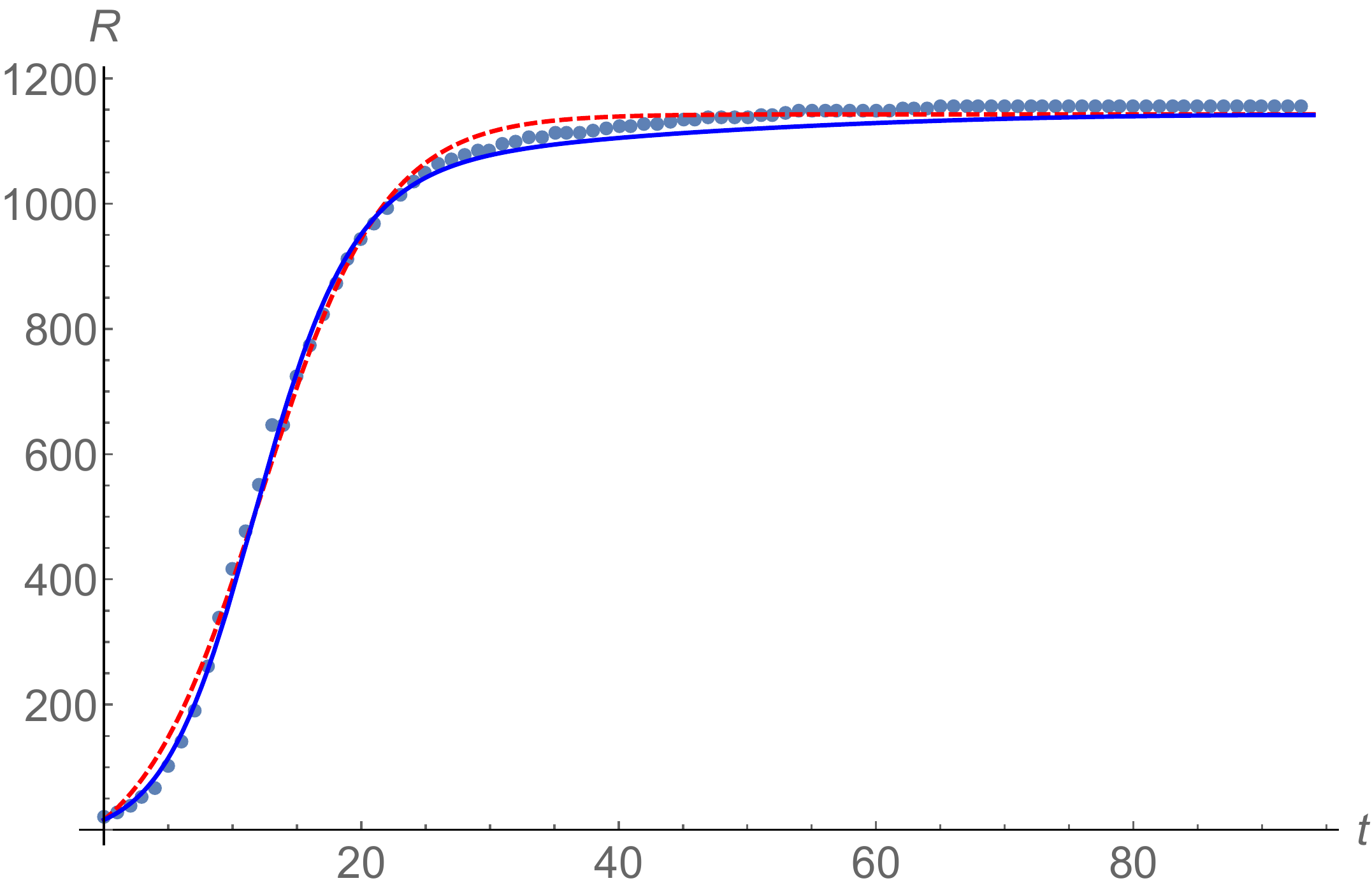} &   \includegraphics[width=80mm]{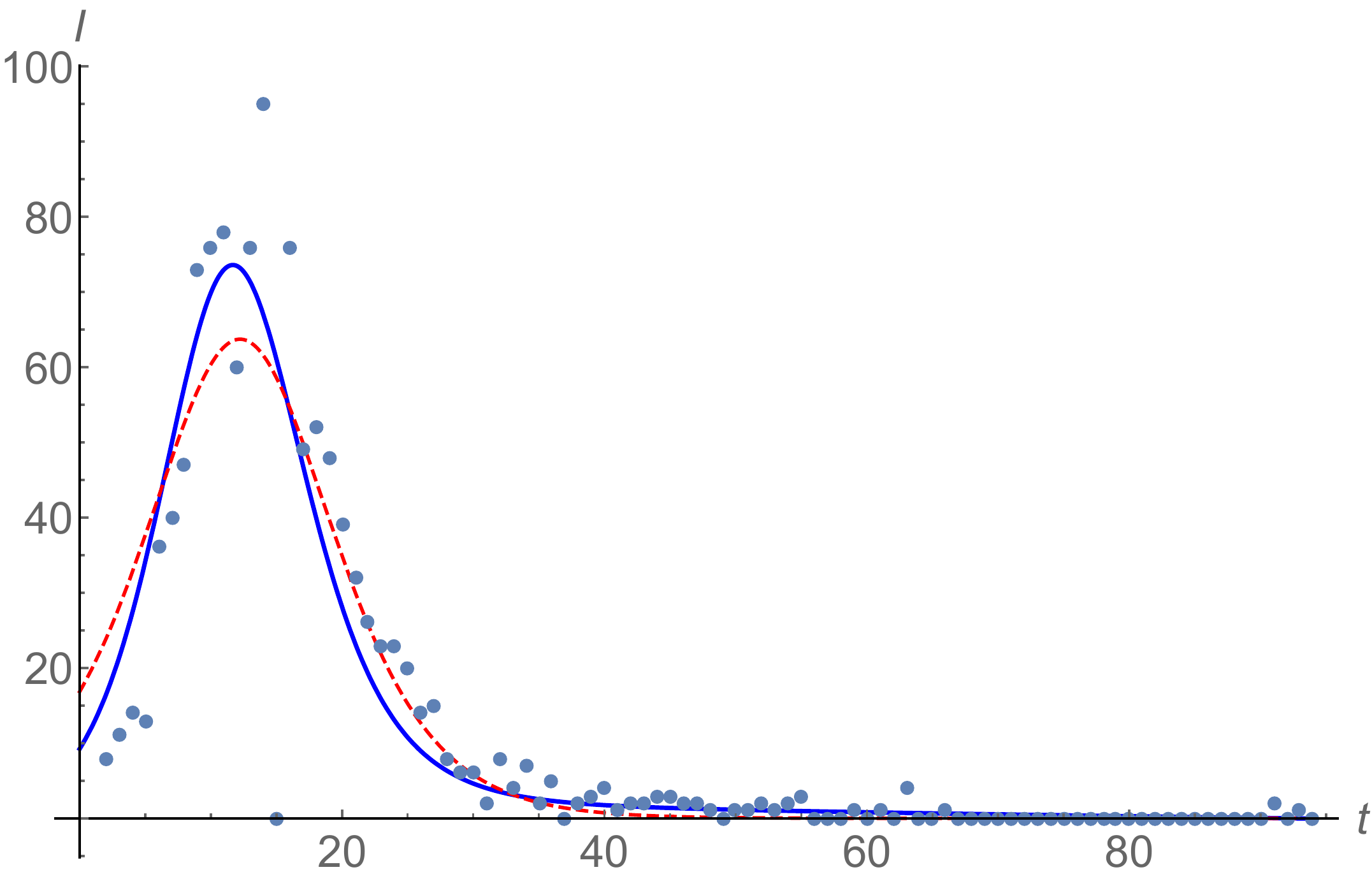} \\
(c) Recovered population for New Zealand & (d)  Infected population for  New Zealand \\[6pt]
\includegraphics[width=80mm]{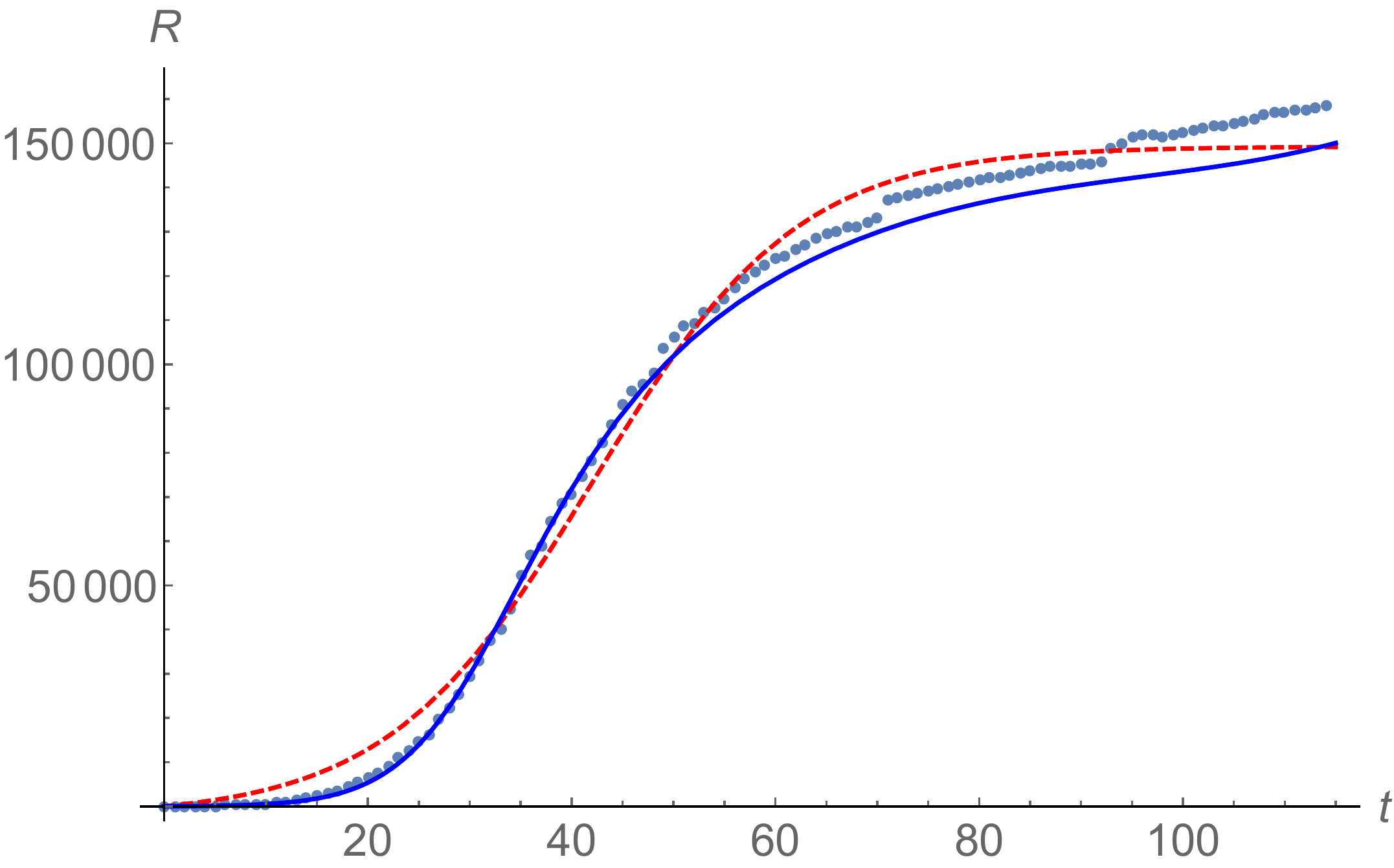} &   \includegraphics[width=80mm]{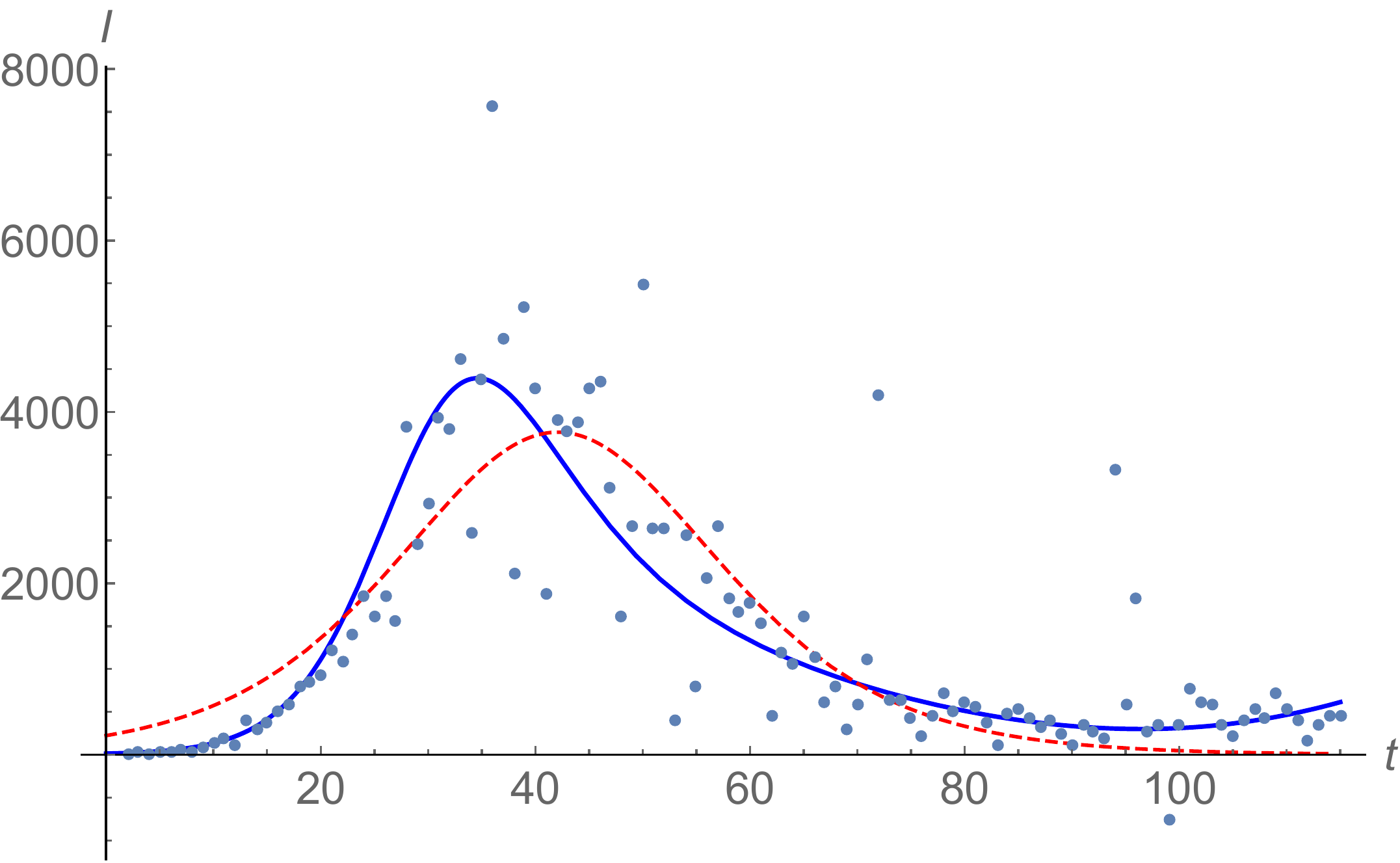} \\
(e) Recovered population for France & (f)  Infected population for France \\[6pt]
\end{tabular}
\caption{Recovered and Infected populations for Portugal, New Zealand and France. Data is shown in dotted line.
Fittings \eqref{Req1b} and \eqref{Ieq1} are shown in red dashed line for recovered and and infected populations, respectively. Phenomenological fittings  \eqref{Req2} and \eqref{Ieq2}, with meta--parameters, are shown in blue solid  line for recovered and and infected populations, respectively.}
\label{datasRI3}
\end{figure*}

The recovered and infected population data of the last country we consider, France, are shown in Figs.~\ref{datasRI3}(e) and (f). The fitting \eqref{Req1b} and \eqref{Ieq1} are in red dashed line with $N=115$ and $R_0=14$, $a=75771.6$, $b= 0.04966$, and $c= -2.09013$.
The data is in dotted lines. 

\begin{figure*}[ht]
\begin{tabular}{c}
  \includegraphics[width=160mm]{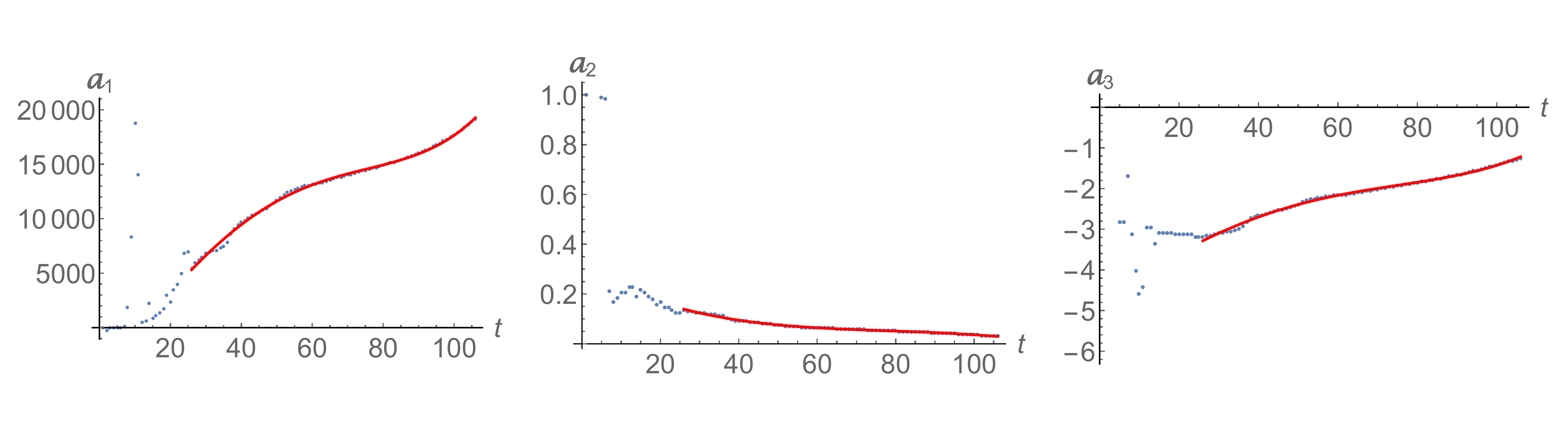} \\
(a) Meta--parameters $a_1$, $a_2$, and $a_3$ for Portugal \\[6pt]
 \includegraphics[width=160mm]{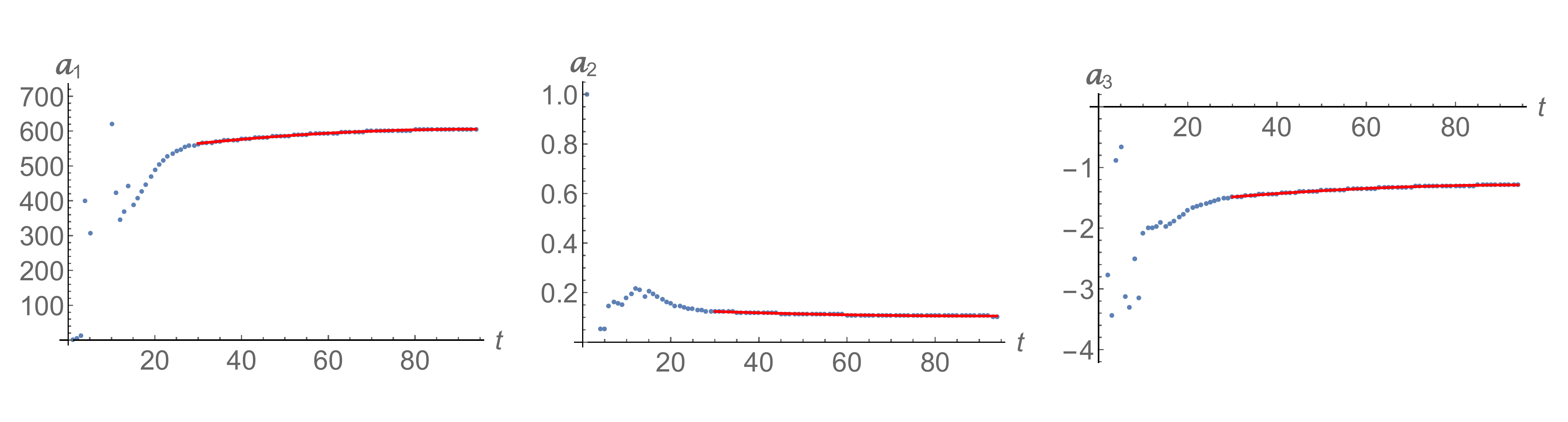} \\
(b) Meta--parameters $a_1$, $a_2$, and $a_3$ for  New Zealand \\[6pt]
\includegraphics[width=160mm]{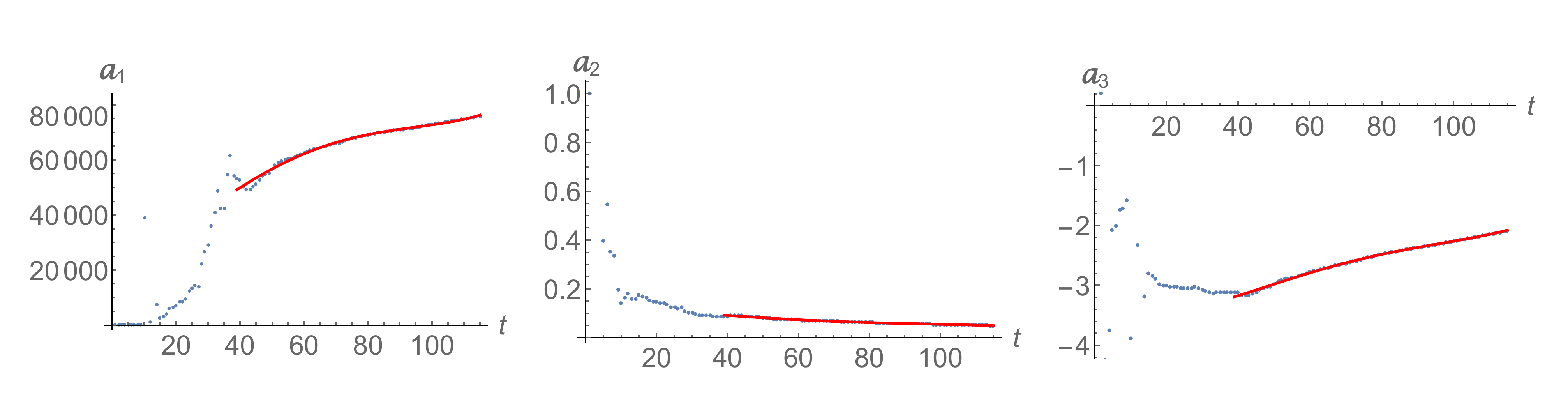} \\
(c) Meta--parameters $a_1$, $a_2$, and $a_3$ for France \\[6pt]
\end{tabular}
\caption{Meta--parameters for Portugal, New Zealand and France. Data is shown in dotted line.
The regular dynamical behavior of each meta--parameter (starting in time $\zeta$) is shown in  red solid  line. They evolve following the quadratic form \eqref{formmetaparametera}.}
\label{datasM3}
\end{figure*}

Blue lines are fitting  \eqref{Req2} and \eqref{Ieq2} with meta--paramters from Figs.~\ref{datasM3}(c), with $Z=4$ and $\tau=39$. Our meta--analysis adjustment gives $\epsilon=0.77846$, and it predicts a slightly increase of the infected population.

\section{Discussion}

We have presented a new approach to deal with infection propagation data by allowing the parameters to became time--dependent. Although we are not able to produce a dynamical model for the parameters time evolution due to intrinsic difficulty associated with unforeseeable government policies and population behavior, we have been able to produce a method that may be successfully applied to the actual data informed by eleven different countries which have implemented different mitigation policies to fight the COVID-19 infection with different population reactions. 

All of the cases of infected populations studied above exhibit the same feature. The meta--analysis shows effectively the capture of the daily variations of cases. In other words, we have shown that using meta--parameters we can {\it integrate} the recovered population without using any pre-existing model. 
Our proposal produces global results, as soon as the regular behavior of meta--parameter are found. Thus, the meta--analysis works for all data, and not only for an arbitrary particular range in the evolution of recovered cases, as for example, when the growing of recovered cases behave as a power-law \cite{manchein}.  This proposed model can also detect sub--epidemic events occurring during the pandemic, which is done without the assumption of any model based in differential equations.

Our model has two main  key characteristics that need to be considered
in order to produce sensible results. The first one is the starting model required to fit the data. In our case, we use a $\tanh$ fit \eqref{Req1b}, but other more complex models, as modified logistic ones, can be used as a starting point for the meta--analysis. This can improve the fitting done by the meta--parameters.
The second one is the high sensitivity of our model to well--established data for cases. The infection data that is informed in countries where the governments have not been  able to handle the pandemic or
 have introduced sudden infection related policy changes, or where the population has not abided by the government set containment rules, are difficult to be described by the approach presented here (or almost any other method). This is because no regularity can be found when the meta--parameters are analyzed. This is the case of, for example,  Chile, where the health authorities policies designed to address the pandemic, and the response of the society, have proved unsuccessful up to now.

Data Availability Statements: 
The data that support the findings of this study are openly available in Complete Our World in Data COVID-19 dataset, at ourworldindata.org.

\begin{acknowledgments}
The authors thank Gianni Tallarita for interesting discussions and help with coding in the initial stages of this work.  F.A.A. thanks Fondecyt-Chile Grant No. 1180139. 
\end{acknowledgments}


\begin{thebibliography}{}
\bibitem{sir}  W. O. Kermack and A. G. McKendrick, Proc. Roy. Soc. A {\bf{115}}, 700 (1927). 
\bibitem{abd} M A. Abdelkader, Math. Biosc., {\bf{29}}, 293 (1974).
\bibitem{bai} N. T. J. Bailey, {\it {The mathematical theory of infectious diseases and its applications.}} Hafner Press [Macmillan Publishing Co., Inc.] New York, second ed., (1975).
\bibitem{het} H. W. Hethcote, SIAM Review {\bf{42}}, 599 (2000).
\bibitem{katia} S. Towers, K. Vogt Geisse, Chia-Chun Tsai, Q. Han, and Z. Feng, Math. Biosc. and Eng. {\bf{9}}, 413 (2012). 
\bibitem{har} T. Harko, F. S.N. Lobo and M.K. Mak, Applied Math. and Comp. {\bf{236}}, 184 (2014). 
\bibitem{boh}  M. Bohner, S. Streipert, and D. F. M. Torres, Nonlinear Analysis: Hybrid Systems {\bf{32}}, 228 (2019). 
\bibitem{wuhan1} J. T. Wu, K. Leung, G. M. Leung, The Lancet {\bf{395}}, 689 (2020).
\bibitem{harko} T. Harko and Man Kwong Mak, arXiv:2006.07170 (2020).

\bibitem{zhang} H. Zhang {\it et al.}, {\it Dynamic Estimation of Epidemiological Parameters of COVID-19 Outbreak and Effects of Interventions on Its Spread},  dx.doi.org/10.2139/ssrn.3566598 (2020).
\bibitem{lorenzo} L. Mangoni and M. Pistilli, {\it Epidemic Analysis of COVID-19 in Italy by Dynamical Modelling}, dx.doi.org/10.2139/ssrn.3567770 (2020). 

\bibitem{Mummert1} A. Mummert and O. M. Otunuga,  J. Math. Biol. {\bf 79}, 705 (2019).
\bibitem{Mummert2} A. Mummert , J. Math. Biol. {\bf 67}, 483 (2013).
\bibitem{capis} M. A. Capistr\'an, M. A. Moreles and B. Lara, Bull Math Biol 71:1890–190  (2009).
\bibitem{Pollicott} M. Pollicott, H. Wang and H. Weiss, J. Biol. Dyn. {\bf 6}, 509 (2012).
\bibitem{hadeler} K. P. Hadeler,  Math. Biosci. {\bf 229}, 185 (2011).
\bibitem{Ungarala} S. Ungarala, K. Miriyala and T. B. Co, 10th IFAC international symposium on dynamics and control of process systems, The international federation of automatic control, India  (2013).
\bibitem{khan} H. Khan, R.  N. Mohapatra, K. Vajravelu and  S. J. Liao, Appl. Math. Comput. {\bf 215}, 653 (2009).
\bibitem{chowell} G. Chowell, A. Tariq and J. M. Hyman, BMC Medicine {\bf 17}, 164  (2019).
\bibitem{chowell2} G. Chowell, L. Sattenspiel, S. Bansal and C. Viboud, Phys. Life Rev. {\bf 18}, 66 (2016).
\bibitem{chowell3} K. Roosa {\it et al.}, J. Clin. Med.  {\bf 9}, 596 (2020).



\bibitem{tesis} D. Barros de Souza, {\it Analytic Solutions to Stochastic Epidemic Models}, Master thesis, Centro de Ci\^encias Exatas e da Natureza, Universidade Federal de Pernambuco (2017).

\bibitem{bailey} N. T. Bailey, {\it et al.}, {\it The mathematical theory of infectious diseases and its applications.} (Charles Griffin \& Company, 1975).


\bibitem{oneworld} M. Roser, H. Ritchie, E. Ortiz-Ospina and J. Hasell, {\it Complete Our World in Data COVID-19 dataset}. Retrieved from: ‘OurWorldInData.org’ [Online Resource] (2019).


\bibitem{note} The different coefficients of the different meta--parameters for each country are not very enlighting to put it along the work. The authors can provide the coefficients and the numerical codes used in this work to any interested reader.
\bibitem{manchein} C. Manchein {\it et al.}, Chaos {\bf  30}, 041102 (2020).

\end{thebibliography}
\end{document}